\documentclass[
a4paper,
aps,
prd,
preprint,
preprintnumbers,
superscriptaddress,
showpacs,
showkeys,
nofootinbib
]{revtex4-1}

\usepackage{fullpage}
\usepackage{amsfonts}
\usepackage{amsmath}
\usepackage{slashed}
\usepackage{amssymb}
\usepackage{graphicx}
\usepackage{epic}
\usepackage{eepic}
\usepackage{epsfig}
\usepackage{latexsym}
\usepackage{color}
\usepackage{float}
\usepackage{multirow}
\usepackage{hyperref}
\usepackage[caption=false]{subfig}
\usepackage{natbib}

\usepackage{shorthand}

\begin{document}

\title{Effect of temperature and magnetic field on two-flavor superconducting quark matter}

\author{Tanumoy Mandal}
\email{tanumoy.mandal@physics.uu.se}
\affiliation{Department of Physics and Astronomy, Uppsala University, Box 516, SE-751 20 Uppsala, Sweden}

\author{Prashanth Jaikumar}
\email{prashanth.jaikumar@csulb.edu}
\affiliation{Department of Physics and Astronomy, California State University Long Beach, 1250 Bellflower Boulevard, Long Beach, California 90840, USA}

\begin{abstract} 
We investigate the effect of turning on temperature for the charge neutral phase of two-flavor color 
superconducting (2SC) dense quark matter in presence of constant external magnetic field. Within the 
Nambu-Jona-Lasinio model, by tuning the diquark coupling strength, we study the interdependent 
evolution of the quark Bardeen-Cooper-Schrieffer gap and dynamical mass as functions of temperature and magnetic field. 
We find that magnetic field $B \gtrsim 0.02$ GeV$^2$ ($10^{18}$~G) leads to anomalous temperature behavior of the 
gap in the gapless 2SC phase (moderately strong coupling), reminiscent of previous results in the 
literature found in the limit of weak coupling without magnetic field. The 2SC gap in the strong 
coupling regime is abruptly quenched at ultrahigh magnetic field due to the mismatched Fermi surfaces of up and down quarks imposed by charge neutrality and oscillation of the gap due to Landau level quantization. The dynamical quark mass also displays strong oscillation and magnetic catalysis at high magnetic field, although the latter effect 
is tempered by nonzero temperature. We discuss the implications for newly born compact stars with superconducting quark cores.
\end{abstract}


\maketitle 


\section{Introduction}
\label{sec:intro}

Quark matter displays color superconductivity/superfluidity at large baryon density 
and low temperature, which has been in the realm of conjecture since the late 
seventies~\cite{Barrois:1977xd,Bailin:1983bm,Iwasaki:1994ij}. A deconfined state of 
quark matter at weak coupling was predicted~\cite{Collins:1974ky} 
shortly after quantum chromodynamics (QCD) was verified as the correct theory of the strong
interaction, and interest in the field of dense quark matter was reignited by the discovery that color superconductivity in quark matter is favored even at strong 
coupling~\cite{Alford:1997zt,Rapp:1997zu,Berges:1998rc,Alford:1998mk,Son:1998uk}. This makes it a viable component of matter in the interior of neutron 
stars, where densities are moderately high in terms of the QCD scale  $\Lambda_{\rm QCD}\sim$ 
0.2 GeV. Compact stars, therefore, could serve as an astrophysical laboratory to test for 
physical phenomena that arise as a result of color superconductivity. 

The first decade of work in this ``reemergent'' era of color superconductivity, which mostly delineated the various possible phases of color superconducting (CSC) matter, is succinctly reviewed in~\cite{Alford:2007xm}. Since then, the focus has shifted to working out transport coefficients~\cite{Manuel:2007pz,Alford:2008pb} and the role of collective modes in color superconductors~\cite{Brauner:2009df,Braby:2009dw,Alford:2014doa} as the field orients itself towards observational consequences of quark cores in neutron stars. It is worth noting that modern equations of state with color superconductivity in the core support neutron stars with maximum mass exceeding the $2M_{\odot}$ lower limit~\cite{Weissenborn:2011qu,Klahn:2013kga,Ferrer:2015vca}. Neutron stars can have an extremely strong magnetic field, with the surface field for magnetars estimated at $10^{-6}-10^{-5}$ GeV$^2$~\cite{Duncan:1992hi,Paczynski:1992zz,Guver:2007ky}.\footnote{In this paper we express magnetic field
in unit of GeV$^2$ for our convenience. The conversion relation between GeV$^2$ and Gauss is $1~\textrm{GeV}^2=5.13\times 10^{19}$~G.} Magnetized phases of color superconducting matter even allow for field values as large as order 1 GeV$^2$ in the dense core~\cite{Ferrer:2010wz}, exceeding the limit set by the virial theorem~\cite{Cardall:2000bs} due to pressure anisotropy effects on the stability of a self-bound star. Such large magnetic fields qualitatively change not only the nonmagnetic Bardeen-Cooper-Schrieffer (BCS) state~\cite{Ferrer:2005vd,Ferrer:2006vw,Ferrer:2007iw,Fukushima:2007fc,Noronha:2007wg}, but also the chiral condensate~\cite{Ferrer:2013noa}, another feature of nonperturbative QCD. 

At the same time as these advances in cold dense quark matter have occurred, there has been
growing interest in the magnetic field of hot quark matter created transiently in heavy-ion
collisions, such as at the Relativistic Heavy Ion Collider and the Large Hadron Collider. It is estimated  that the
magnetic field generated from the chiral magnetic effect in off-central nucleon-nucleon collisions at these colliders can 
be as large as $0.01-1$ GeV$^2$~\cite{Kharzeev:2007jp,Skokov:2009qp,Voronyuk:2011jd}. Conversely, there are several studies of magnetic catalysis of chiral symmetry breaking at zero density and temperature~\cite{Gusynin:1994re,Semenoff:1998bk,Ferrer:1999gs,Kabat:2002er,Miransky:2002rp} (for more references see {\it e.g.}~\cite{Shovkovy:2012zn} and references therein), as well as restoration of chiral symmetry/inverse magnetic catalysis at high 
temperature~\cite{Avancini:2011zz,Fukushima:2012kc,Bali:2013esa,Mueller:2015fka,Ahmad:2016iez} or with increasing quark chemical potential~\cite{Gorbar:2009bm,Preis:2010cq,Chatterjee:2011yi}.  
The wide-ranging physics of the magnetic field in strongly interacting relativistic and 
nonrelativistic forms of matter is reviewed in~\cite{Kharzeev:2013jha}. Our work is an addition to the literature aimed at determining the evolution of chiral and diquark condensates and ultimately the phase structure of magnetized dense quark matter as functions of external controls such as the magnetic field and coupling strengths. The new considerations in this work, since we are focused on the quark cores of neutron stars, is that of the temperature effect on magnetized quark matter that is color and charge neutral. Specifically, we study the two-flavor color superconducting (2SC) phase of cold and warm dense quark matter, ignoring the role of the strange quark for the purposes of this paper. However, it can be included along the lines discussed in~\cite{Mishra:2004gw}, which studied gapless modes at finite temperature for 2SC+s matter.
In Refs.~\cite{Ruester:2005jc,Blaschke:2005uj}, it was shown that the 2SC phase is favored as the ground state of
CSC matter in the strong diquark-to-scalar coupling ratio ({\it i.e.}, $G_D/G_S \sim 1$). 
The couplings $G_D$ and $G_S$ appear later in Eq.~\eqref{eq:lag}.

In a previous paper~\cite{Mandal:2012fq}, we discussed the effect of imposing color and charge neutrality conditions on the 2SC
phase at zero temperature. Solving the gap equations for the evolution of the condensates, we found that in the charge neutral gapless 2SC (g2SC) phase~\cite{Huang:2003xd}, which occurs at moderately large diquark coupling, a large magnetic field drives the CSC phase transition to a crossover, while the chiral phase transition is first order. The charge neutrality condition, 
in particular, leads to an additional stress on pairing due to the Fermi surface mismatch for two flavors, leading to smaller values of the gap than for the non-neutral case. For the very strong field, 
$B\gtrsim 0.4$~GeV$^2$, even at strong coupling, the homogeneous 2SC gap can vanish due to de Hass-van Alphen oscillations that disrupt pairing~\cite{Mandal:2012fq}. In this work, we extend our studies to finite temperature up to tens of MeV, in order to describe a wider range of physical conditions under which quark matter can form, and color superconductivity possibly suppressed, in (proto)neutron stars. An additional reason to study the effect of temperature is that it can lead to a reinforcement of pairing in asymmetric systems, and suppression of the chiral condensate due to thermal fluctuations. This can qualitatively change the results from those at zero temperature.

A study of the 2SC phase at finite temperature and high magnetic field was conducted in~\cite{Fayazbakhsh:2010bh}; however, neutrality was not imposed nor was the competition between condensates allowed for as a function of the couplings, so that the results cannot be applied directly to compact stars. By including neutrality, we arrive at qualitatively new results, such as anomalous behavior of the gap with temperature and abrupt quenching of the gap at critical value of the magnetic field depending on the temperature window. These findings could be relevant to the temperature and density regime of protoneutron stars with quark matter at temperature below the CSC phase transition.

The paper is organized as follows: In Sec.~\ref{sec:lag}, we state the model Lagrangian and its parameters, recast the partition function in terms of interpolating bosonic variables and derive the thermodynamic potential. Extremization then provides the equations that yield the relevant chiral/diquark gaps as well as the neutralizing charges, which must be solved for numerically at nonzero temperature. In Sec.~\ref{sec:numana}, we discuss the physics behind the numerical results obtained for the coupled evolution of the condensates as a function of density, magnetic field, and temperature for various choices of the diquark-to-scalar coupling ratio. We summarize our main findings in Sec.~\ref{sec:conc}.


\section{Lagrangian and Thermodynamics}
\label{sec:lag}

The formalism we employ is a straightforward extension to finite temperature of the model 
used in~\cite{Mandal:2012fq} at zero temperature, hence we do not repeat all the details, 
merely the essential parts. We employ a Nambu-Jona-Lasinio (NJL)-type Lagrangian invariant under global 
$\mathrm{SU(2)_L \times SU(2)_R}$ in the massless quark limit, while the diquark condensate 
breaks $\mathrm{SU(3)_c\to SU(2)_c}$. The Lagrangian density reads as
\begin{eqnarray}
\label{eq:lag}
\mathcal{L} &=& \bar{q}\left[i\gamma^{\mu}\left(\partial_{\mu} - ieQA_{\mu} - igT^8G^8_{\mu}\right) 
+ \hat{\mu}\gamma^{0} - \hat{m}\right]q + G_{S}\left[\left(\bar{q}q\right)^2
+ \left(\bar{q}i\gamma_{5}\vec{\tau}q\right)^2\right]\nonumber\\
&+& ~G_D\left[\left(\bar{q}i\gamma_{5}\epsilon_{f}\epsilon_{c}q^C\right)
\left(\bar{q}^{C}i\gamma_{5}\epsilon_{f}\epsilon_{c}q\right)\right]\ ,
\end{eqnarray}
with quark spinor fields $q\equiv q_{ia}$ indexed by $i=(1,2)=(u,d)$ for the flavor doublet and $a=(1,2,3)=(r,g,b)$ 
for the color triplet; $q^C=C\bar{q}^T$ and $\bar{q}^C=-q^TC$ are the charge-conjugated fields of $q$ and 
$\bar{q}$, respectively with $C=-i\gamma^0\gamma^2$.
The chemical potential for any flavor and color is given by $\hat{\mu}=\mu - Q\mu_e + T^3\mu_{3c} + T^8\mu_{8c}$,
where $\mu$ is the common quark chemical potential. 
The current quark mass matrix in flavor basis is $\hat{m}\equiv \textrm{diag}(m_u,m_d)$. We assume the $\mathrm{SU(2)}$ isospin 
symmetry is exact i.e., $m_u=m_d=m_0$ where $m_0$ is the current quark mass. Finally, the coupling strengths in the scalar and the diquark channels are denoted by $G_S$ and $G_D$, respectively. After bosonization, we obtain
\begin{eqnarray}
\mathcal{L} &=& \bar{q}\left[i\gamma^{\mu}\left(\partial_{\mu} - ieQA_{\mu} - igT^8G^8_{\mu}\right) + \hat{\mu}\gamma^{0}\right]q 
- \bar{q}\left(m + i\gamma_5\vec{\pi}\cdot\vec{\tau}\right)q \nonumber\\
&-& \frac{1}{2}\Delta^{*}\left(\bar{q}^{C}i\gamma_{5}\epsilon_{f}\epsilon_{c}q\right) - \frac{1}{2}\Delta\left(\bar{q}i\gamma_{5}\epsilon_{f}\epsilon_{c}q^C\right)
- \frac{\sigma^2 +
\vec{\pi}^2}{4G_s} - \frac{\Delta^{*}\Delta}{4G_D}\ ,
\end{eqnarray}
where $m = m_0 + \sigma$. Strictly speaking, the emergent constituent mass $m$ can be flavor dependent since isospin symmetry is explicitly broken by a nonzero $\delta\mu=\mu_d-\mu_u$ and magnetic field $B$, but we ignore this effect here for simplicity. Taking this effect into account increases the number of coupled gap equations to solve, and will be examined in future work.
The vector $\vec{\pi}=0$ implies that we disregard the possibility of pion condensation~\cite{Andersen:2007qv}. 
Nonvanishing vacuum expectation values (VEVs) for $\sigma$ and $\Delta$ represent chiral symmetry breaking and color superconductivity in quark matter. 
For nonzero $\Dl$, as a result of diquark condensate carrying a net electromagnetic charge, there is a Meissner effect for ordinary magnetism. But a residual $\mathrm{U(1)}$ symmetry still leads to a ``rotated'' massless photon. The rotated combination is $\tl{A}_\mu=A_\mu\cos\theta-G_\mu^8\sin\theta$ where the mixing angle $\theta$ between flavor and color hypercharge 
is given by $\sin\theta = -e/\sqrt{3g^2+e^2}$. The rotated charge matrix in the
$flavor\otimes color$ space in the unit of the rotated charge of an electron 
$\tilde{e}={\sqrt{3}ge}/{\sqrt{3g^2+e^2}}$ is
\begin{eqnarray}
\tilde{Q} = Q_f\otimes{\bf 1}_c - {\bf 1}_f\otimes \frac{T^8_c}{2\sqrt{3}}\ .
\end{eqnarray}
The absence of the other Casimir operator $T^3$ follows from the degeneracy of colors $1$ and $2$,
which ensures that there is no long range gluon $3$-field ($\mu_{3c}=0$).
The rotated $\tilde{Q}$ charges of different quarks for the 2SC phase are presented in Table~\ref{rotcharge}. 
\begin{table}
\begin{center}
\scalebox{1}{
\begin{tabular}{|c|c|c|c|c|c|c|} \hline
\multicolumn{1}{|c|}{Flavor} & \multicolumn{3}{|c|}{up} & \multicolumn{3}{|c|}{down} \\ \cline{1-7} 
\multicolumn{1}{|c|}{Color} & \multicolumn{1}{|c|}{Red} & \multicolumn{1}{c|}{Green} & \multicolumn{1}{c|}{Blue} & \multicolumn{1}{|c|}{Red} & \multicolumn{1}{c|}{Green} & \multicolumn{1}{c|}{Blue}\\ \hline
$\tilde{Q}$ charge  & $+\frac{1}{2}$  & $+\frac{1}{2}$ & 1 & $-\frac{1}{2}$ & $-\frac{1}{2}$ & 0   \\ \cline{1-7} 
\end{tabular}}
\caption{\label{rotcharge}$\tilde{Q}$ charges of quarks in the 2SC phase in units of $\tilde{e}$ in the presence of external rotated magnetic field $\tilde{\bf B}$.}
\end{center}
\end{table}
The difference of chemical potentials between $u$ and $d$ quarks is compensated
by $\mu_e\neq 0$ in the medium. For calculational simplicity, we define the mean chemical potential $\bar{\mu}$
and the difference of the chemical potential $\delta\mu$ as 
\begin{align}
\label{eq:mismatch}
\bar{\mu} &=\frac{1}{2}(\mu_{u_r}+\mu_{d_g})=\frac{1}{2}(\mu_{u_g}+\mu_{d_r})
=\mu - \frac{1}{6}\mu_e + \frac{1}{3}\mu_{8c}\ ,\\
\delta\mu &=\frac{1}{2}(\mu_{d_g}-\mu_{u_r})=\frac{1}{2}(\mu_{d_r}-\mu_{u_g})
=\frac{1}{2}\mu_e\ .
\end{align}

To study nonzero temperature effects ($T\neq 0$), we derive the partition function
from the Lagrangian (with the magnetic field term) using the imaginary time path integral formalism
\begin{eqnarray}
\mathcal{Z} = N \int [d\bar{q}][dq] \textrm{exp} \left\{\int_{0}^{\beta}d\tau\int
d^{3}\vec{x}\left(\tilde{\mathcal{L}} - \frac{B^2}{2}\right)\right\}\ ,
\end{eqnarray}
with $N$ being the normalization constant, $\tilde{\mathcal{L}}$ the
Lagrangian density with in-medium couplings, and $\beta=1/T$.
Decomposing the partition function as 
\begin{eqnarray}
\mathcal{Z} = \mathcal{Z}_c\mathcal{Z}_{u_b d_b}\mathcal{Z}_{u_r d_g, u_g d_r}\ ,
\end{eqnarray}
where $Z_c$ is a multiplicative constant from bosonization, we can write the relevant parts of the thermodynamic potential $\Omega = - T\ln Z/V$ as
\begin{eqnarray}
\ln \mathcal{Z}_{u_b d_b} = \frac{1}{2}\ln\left\{\textrm{Det}(\beta\mathcal{G}_0^{-1})\right\};~~
\ln \mathcal{Z}_{u_r d_g, u_g d_r} = \frac{1}{2}\ln\left\{\textrm{Det}(\beta\mathcal{G}_{\Delta}^{-1})\right\}\ , 
\end{eqnarray}
where the normal ($\mathcal{G}_0^{-1}$) and anomalous ($\mathcal{G}_{\Delta}^{-1}$) propagators are given in~\cite{Mandal:2012fq}. Subsequent simplification for numerical purposes is achieved by unraveling the color-flavor structure of $\tilde{Q}$~\cite{Noronha:2007wg}, introducing energy projectors~\cite{Huang:2001yw} and moving from position to momentum space to facilitate the Matsubara sum, leading to
\begin{eqnarray}
\label{omegaB}
\Omega_{B} &=& \frac{m^{2}}{4G_{S}}+\frac{\Delta^{2}}{4G_{D}} + \Omega_{e}
+ \sum_{a} \Omega_{a}~~\textrm{where}~~a\in 0,1,\frac{1}{2}\ , \\
\Omega_{e}&=& - \left(\frac{\mu^{4}_e}{12\pi^2} + \frac{\mu^{2}_e T^2}{6} 
+ \frac{7\pi^2 T^4}{180}\right)\ ,
\end{eqnarray}
and $\Omega_a$ are the contributions to the $\Omega_{B}$ from the quarks of 
$\tilde{Q}$ charge $a$ (explicit expressions are given in~\cite{Mandal:2012fq}). Two gap equations and two neutrality conditions at finite temperature are obtained from extremization,
\begin{equation}
\frac{\partial\Omega_{B}}{\partial\varsigma}=0\,;\quad \varsigma = \sigma\,,\Delta\,,\mu_{8c}\,,\mu_e\ .
\end{equation}


\section{Numerical analysis and Results}
\label{sec:numana}

The chiral and the diquark gap equations together with the electric and the color charge 
neutrality conditions from the previous Sec.~\ref{sec:lag} are numerically solved to 
describe the evolution of $m$, $\Delta$, $\mu_e$ and $\mu_{8c}$ in matter with nonzero 
temperature and magnetic field. These four coupled equations involve diverging momentum 
integrals in the ultraviolet, requiring regularization through a choice of common cutoff 
schemes~\cite{Fukushima:2007fc,Fayazbakhsh:2010gc,Frasca:2011zn,Allen:2015paa}. While a sharp regulator
(step function) is often easy to implement, this can sometimes lead to unphysical behavior in thermodynamical quantities of interest, especially when dealing with a system of discrete Landau levels. 
Therefore, we use the following Fermi-Dirac-type smooth cutoff function in our analysis,
\be
\label{eq:reguFD} 
f_c(p_a) = \frac{1}{2}\lt[ 1 - \tanh\lt( \frac{p_a - \Lm}{\al}\rt)\rt] \ ,
\ee
where $p_a = \sqrt{{\bf p}_{\perp,a}^2 + p_z^2}$ with ${\bf p}_{\perp,a=0}^2 = p_x^2 + p_y^2$, ${\bf p}_{\perp, a\neq 0}^2=2|a|\tl{e}Bn$ for $a = 1,\pm 1/2$.
Here, $\Lm$ is the cutoff scale and $\al$ is a smoothness parameter, chosen to be 
$\al = 0.01\Lm$ for our numerical analysis. Doubly degenerate Landau levels are labeled by $n$. Our main results are almost insensitive 
to different cutoff schemes. We fix the cutoff scale $\Lambda$ and other free parameters of the NJL model,  
\begin{eqnarray}
\label{eq:njlparam}
\Lambda = 0.6533~\textrm{GeV},~~G_S = 5.0163~{\textrm{GeV}}^{-2},~\textrm{and}~~m_0=0.0055~\textrm{GeV}\ ,
\end{eqnarray}
by fitting to three vacuum quantities, namely, the constituent quark mass ($0.33$ GeV), the pion mass ($0.135$ GeV), and the pion decay constant ($0.0923$ GeV). 
In works such as~\cite{Huang:2002zd}, the current quark mass $m_0=0$ whereas we keep a flavor-independent value $m_0=0.0055$
throughout our analysis, which is reasonable for two light quark flavors. The diquark strength $G_D=\rho G_S$ is determined once the parameter $\rho$ is fixed to a typical value between $0.6$ and $1.15$, which includes the value $\rho=0.75$ suggested by 
Fierz transformation of one-gluon exchange effective four-quark interaction for $N_c=3$ and the value $\rho=2.26/3$~\cite{Ebert:1991pz} from fits to baryon masses. 
As the underlying QCD interaction at moderate density, high magnetic field and temperature
is bound to be more complicated, we choose to vary $\rho$ between $0.6$ and $1.15$ to investigate the competition between the condensates and related critical phenomena.

In studies of the 2SC phase at zero magnetic field~\cite{Huang:2003xd}, the
phase structure of electric and color charge neutral two-flavor quark matter
is found to be very sensitive to $\rho$. The 2SC phase
(i.e., $\Delta > \delta\mu$ for all momentum modes) is found for $\rho\gtrsim 0.8$, while gapless modes appear in a window $0.7 \lesssim\rho\lesssim 0.8$ and normal
quark matter (i.e., $\Delta = 0$) emerges for $\rho \lesssim 0.7$~\cite{Huang:2001yw,Huang:2003xd,Mandal:2009uk}. Turning on a strong magnetic field in neutral matter, our earlier work~\cite{Mandal:2012fq} showed that the CSC phase transition becomes a crossover, and the magnetic field plays an essential role in delineating the breakdown of the homogeneous pairing ansatz. Here, we explore consequences of nonzero temperature on the competition between the chirally broken and diquark phase in the presence of a strong magnetic field. To reiterate, such physical conditions may be realized in the cores of neutron stars when superconducting quark matter is favored.
 
One additional point worth mentioning here is that our results for the weak magnetic field 
($\tl{e}B\lesssim 0.01$ GeV$^2$) are almost identical to zero field results since the discreteness of the energy levels and its effect on the cutoff scale $\Lambda$
is imperceptible in the weak field limit where $n_{max}$, the maximum number of completely occupied Landau levels, becomes very large and the summation over discrete levels becomes 
quasicontinuous. Quantitatively we find that 
magnetic field $\sim 0.01$ GeV$^2$ corresponds to $n_{max}\sim 20$. Therefore, the results 
for $\tl{e}B\lesssim 0.01$ GeV$^2$ become almost identical to zero field results and we choose
$\tl{e}B=0.005$ GeV$^2$ to present our results in the weak field limit. The effect of the magnetic field
becomes clearly visible for $\tl{e}B\gtrsim 0.02$ GeV$^2$ (equivalent to $n_{max} \lesssim 10$) and
we refer it as the strong field regime.

\begin{figure}[!ht]
\subfloat[]{\includegraphics[scale=0.6]{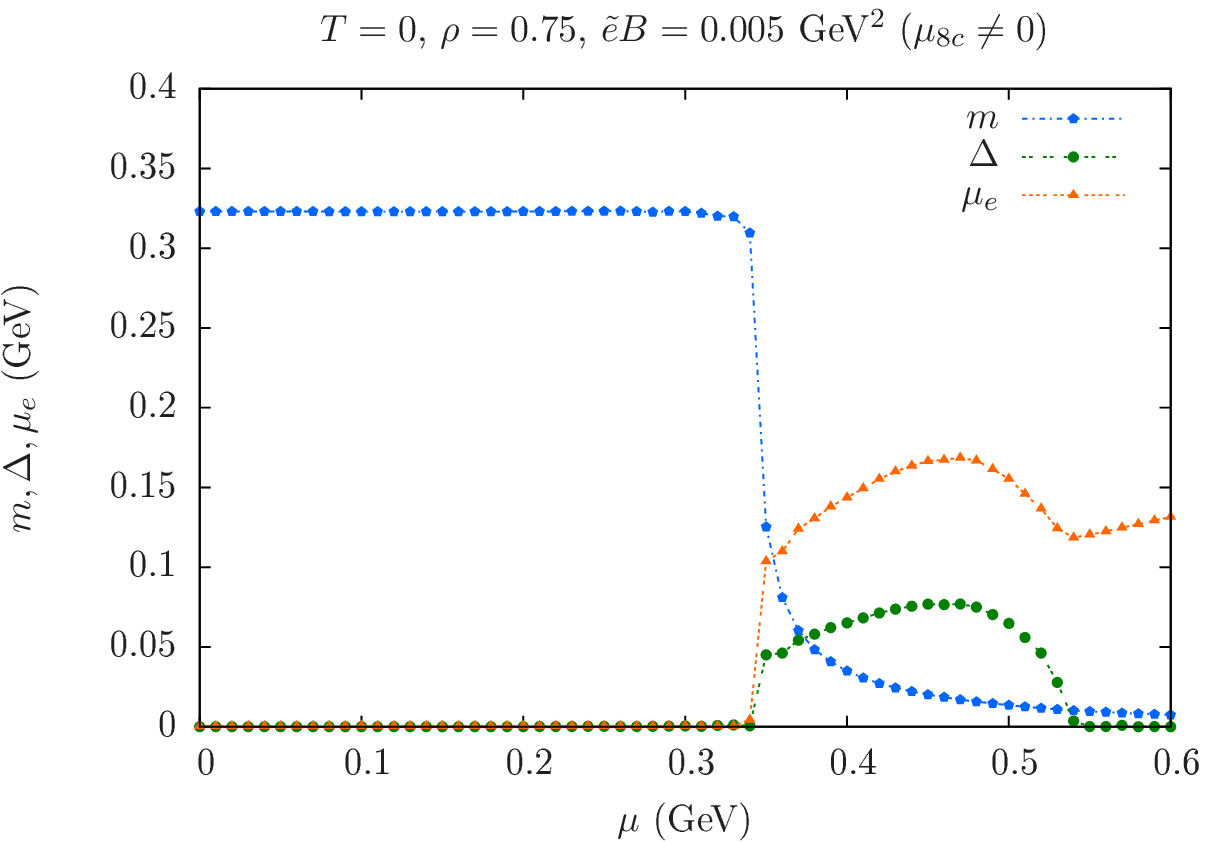}\label{f1a}}
\subfloat[]{\includegraphics[scale=0.6]{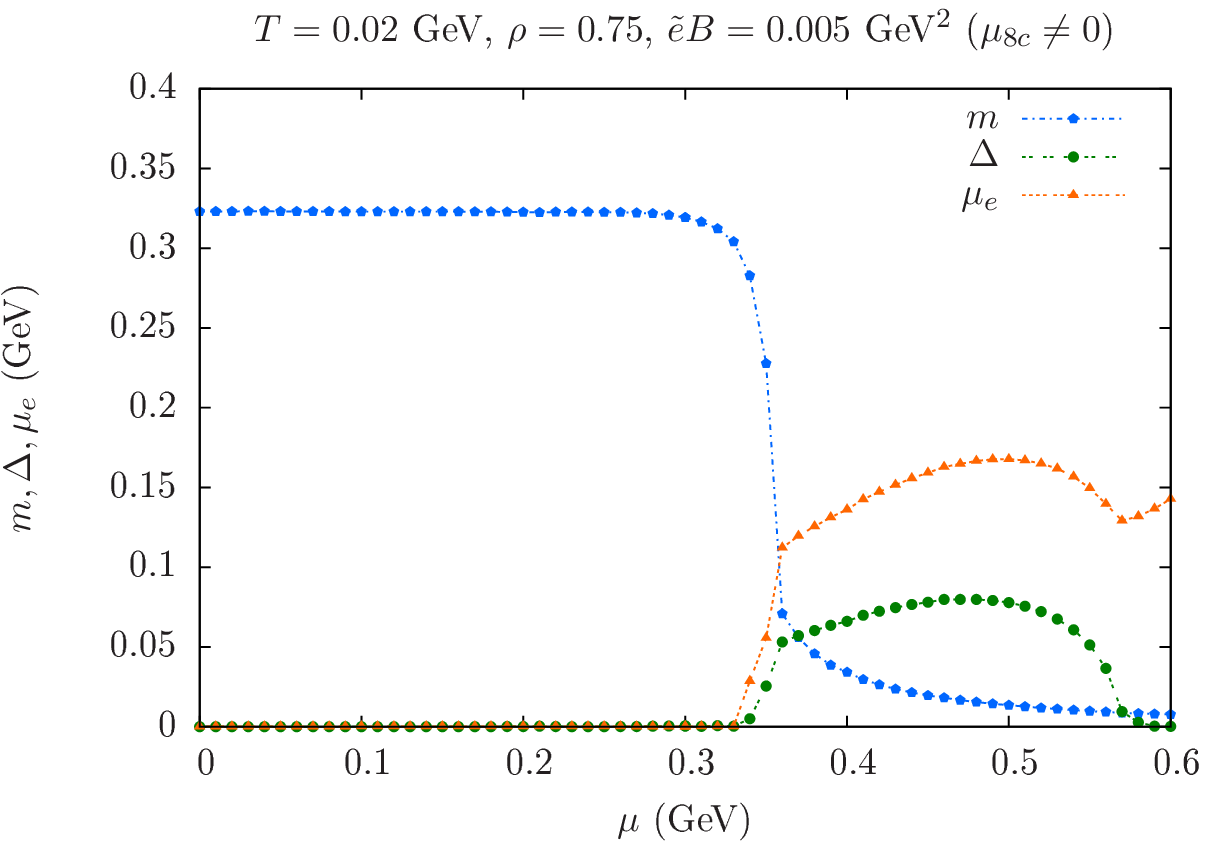}\label{f1b}}\\
\subfloat[]{\includegraphics[scale=0.6]{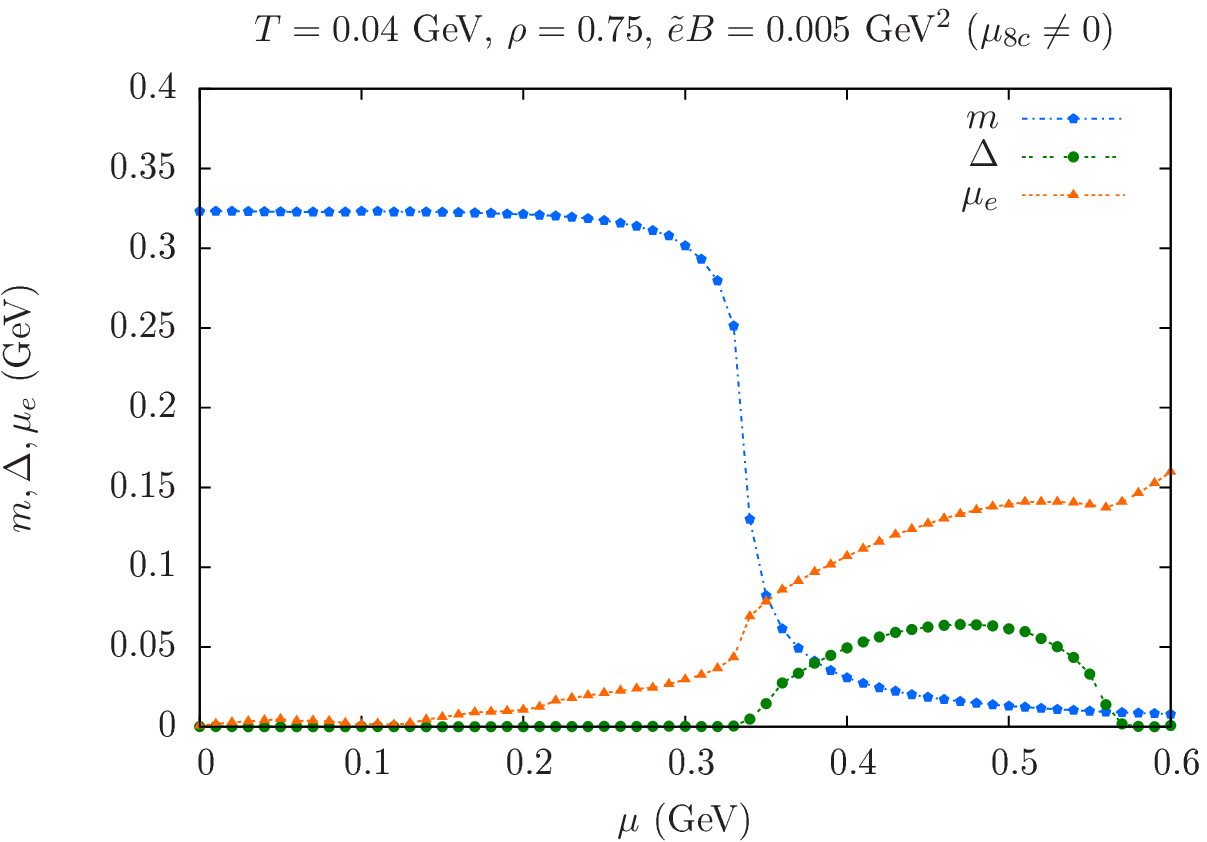}\label{f1c}}
\subfloat[]{\includegraphics[scale=0.6]{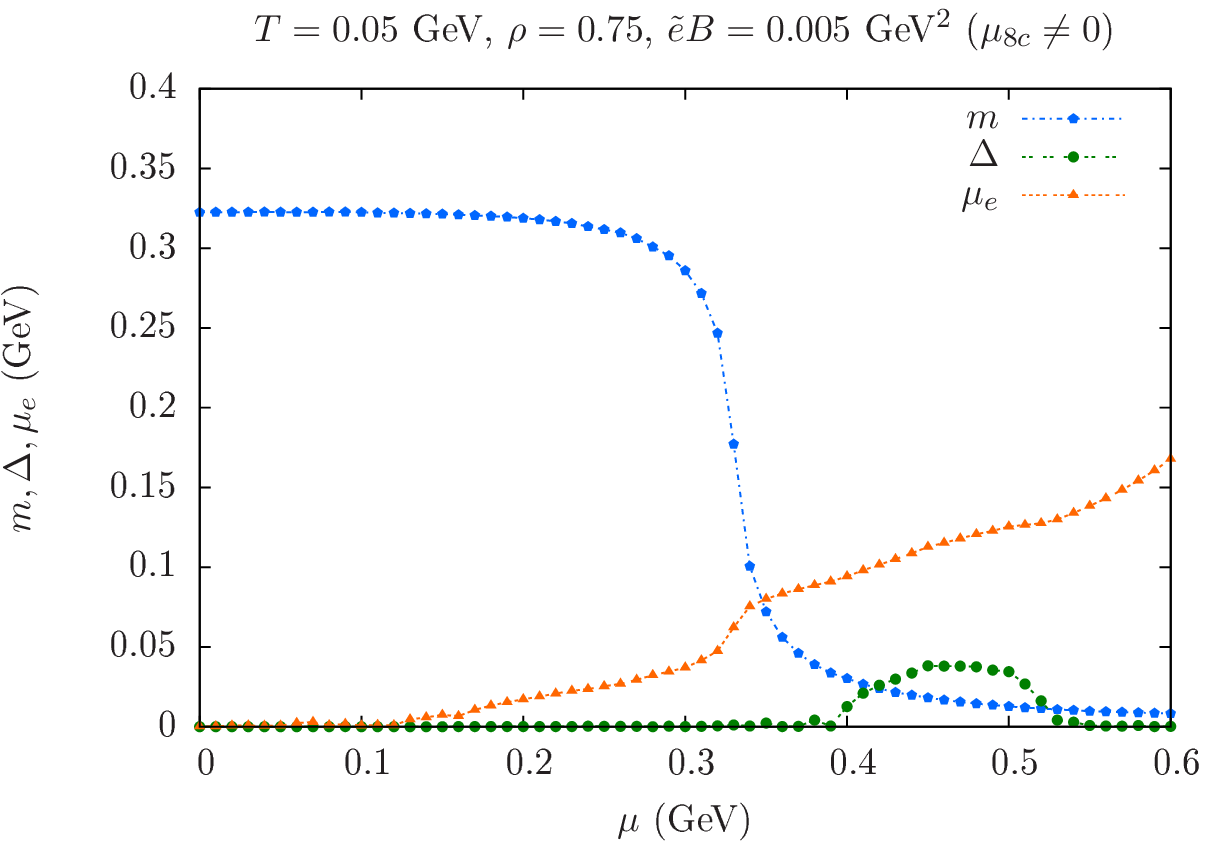}\label{f1d}}
\caption{The dependence of $m$, $\Delta$, and $\mu_e$ on $\mu$ for 
electric and color charge neutral quark matter in the weak magnetic field limit
for $\rho=0.75$. The panel shows four plots for $T=0,0.02,0.04$, and 0.05 GeV.}
\label{fig:1}
\end{figure}

In Fig.~\ref{fig:1}'s panel, we show $m$, $\Delta$, and $\mu_e$ as functions of 
$\mu$ for charge neutral two-flavor quark matter in the weak magnetic field limit 
($\tl{e}B=0.005$ GeV$^2$) for increasing temperature. The relevant range of density for the cores of neutron stars is 
$\mu\sim (0.3-0.4)$ GeV while the temperature of protoneutron stars can be as large as 0.05 GeV. The choice $\rho=0.75$ for all figures in this panel corresponds 
to the g2SC phase. Since we are not in the chiral limit, the constituent mass does not 
vanish at any $T$ or $\mu$, and 
the chiral transition is first order at a pseudocritical chemical potential $\mu_c\approx m(\mu)=0.315$ GeV, 
while the CSC gap appears at the same or higher $\mu$ depending on the temperature.
We find that the order of chiral phase transition is first order in nature as the first derivative of the free energy with respect to the chiral field is discontinuous. The results for the weak magnetic field limit shown in Fig.~\ref{f1a} are in very good agreement with the zero field results in Ref.~\cite{Huang:2002zd}, which is expected since the number of 
completely populated Landau levels is large. It is important to mention that the smooth decrease in the dynamical quark mass after $\mu_c$ is an artifact of using a realistic nonzero current quark mass $m_0=0.0055$ GeV. As a consequence of this, explicit chiral symmetry breaking is alive even at large density.
After the first order transition at $\mu_c$, chiral symmetry is only partially restored.
In the g2SC phase, $\mu_e/2=\dl\mu>\Dl$, which is confirmed in all plots in Fig.~\ref{fig:1}. As the temperature is increased, from Figs.~\ref{f1a} - \ref{f1d}, the magnitude of the CSC gap first increases slightly up to $T\approx 0.02$ GeV, then decreases. This nonmonotonic behavior with temperature for the g2SC phase is a result of thermal motion, which first helps to bridge the mismatch in the $u$ and $d$ Fermi surfaces of red and green colors to enhance pairing, then disrupts it~\cite{Huang:2003xd}. Furthermore, the first order chiral transition at $\mu_c\sim m(\mu)$ becomes smoother due to the smearing of the Fermi surface by temperature (for all colors). At zero temperature, $\mu_e$ is zero in the neutral, chirally broken phase (zero occupation number for quarks), while at finite temperature, $\mu_e\neq 0$ is allowed below the chiral transition. As the occupation numbers build up, the difference in quasiparticle density $(n_{d,g}-n_{u,r})$ increases due to the mismatch $\delta\mu$, and $\mu_e$ increases. Note that while the magnitude of the gap does affect the value of $\mu_e$, the charge imbalance between $u$ and $d$ quarks mainly determines the trend in $\mu_e$.

\begin{figure}[!ht]
\subfloat[]{\includegraphics[scale=0.6]{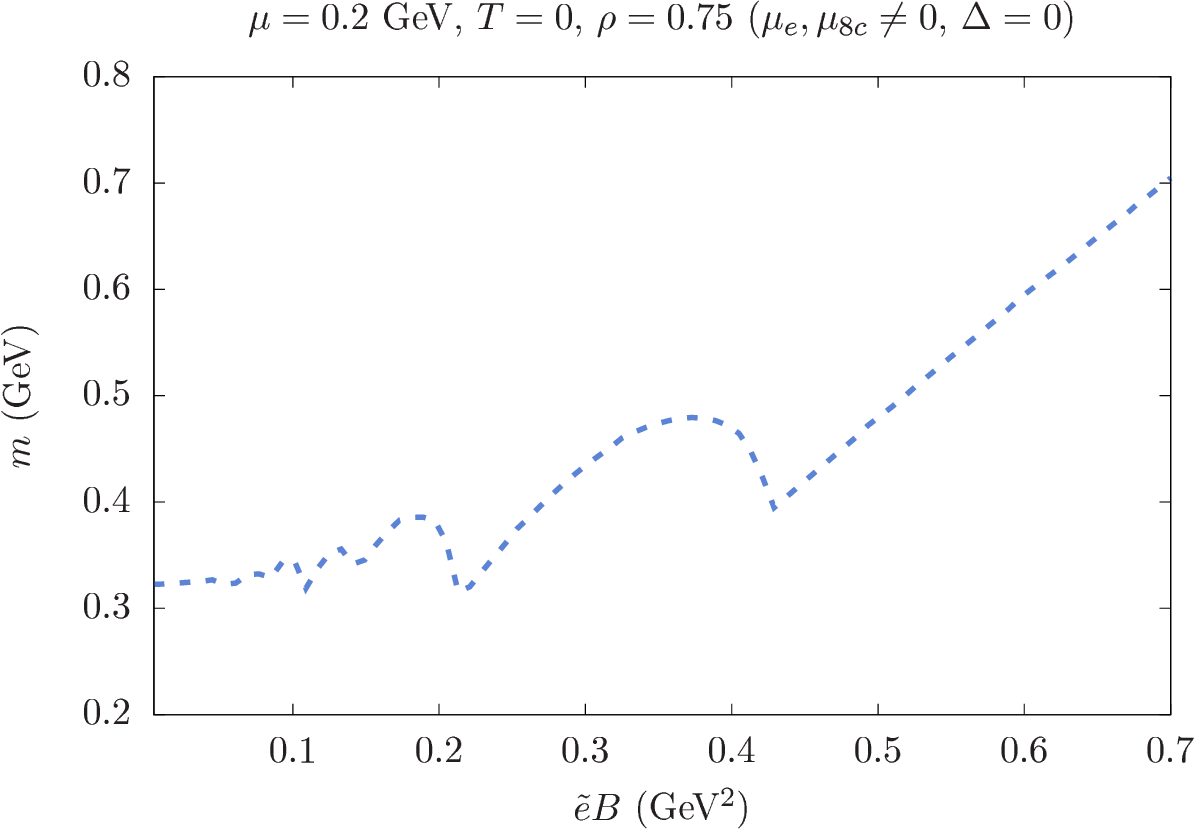}\label{magcat}}
\subfloat[]{\includegraphics[scale=0.6]{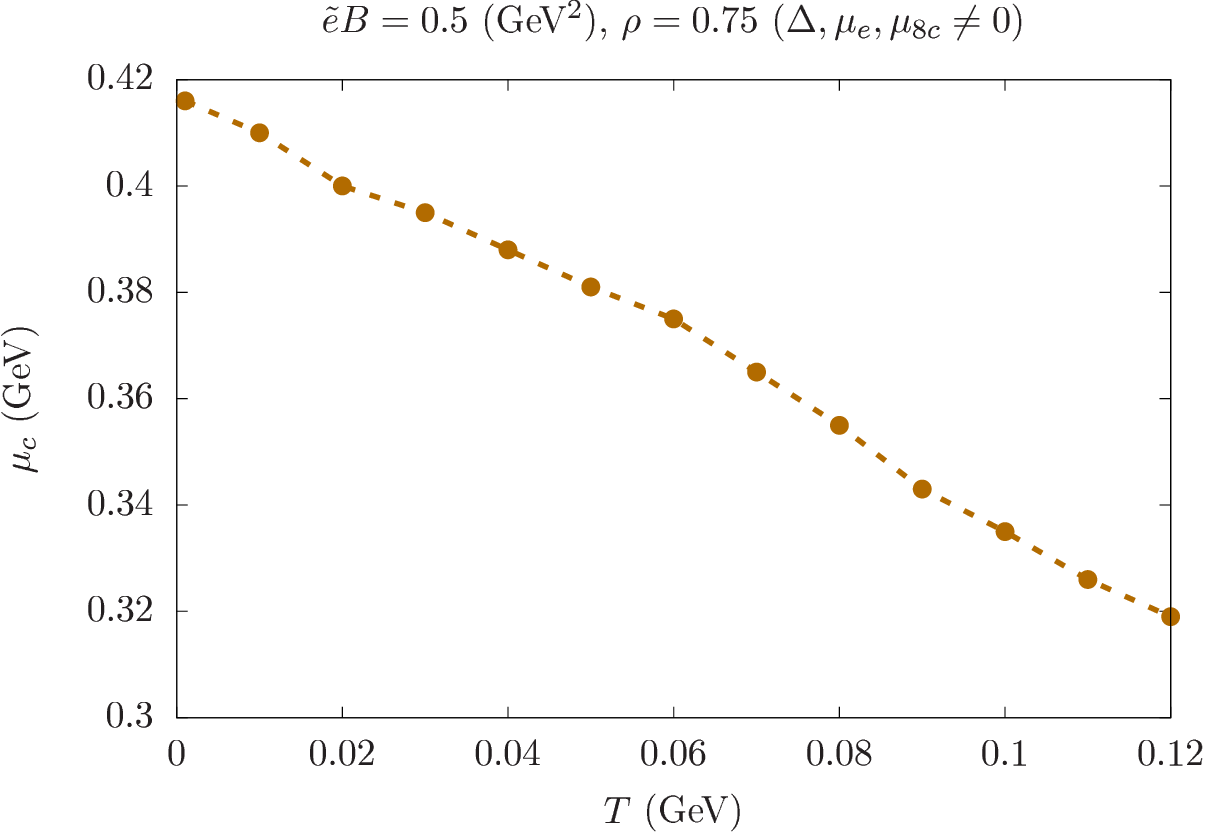}\label{invmagcat}}
\caption{(a) The constituent quark mass $m$ as a function of $\tl{e}B$ at $\mu=0.2$ GeV in the chirally broken phase.
(b) The critical chemical potential ($\mu_c$) for chiral phase transition as a function of $T$ for $\tl{e}B=0.5$ GeV$^2$. For both plot we choose $\rho=0.75$.}
\label{fig:cat}
\end{figure}

We also observe (as shown in Fig.~\ref{magcat}) oscillation of the constituent quark mass ($m$) at higher field values ($\tl{e}B \gtrsim 0.05$ GeV$^2$) and eventually magnetic catalysis of chiral symmetry breaking~\cite{Shovkovy:2012zn} at very large magnetic field $\tl{e}B \gtrsim 0.41$ GeV$^2$ when only the lowest Landau level is occupied. Magnetic catalysis is moderated by increasing temperature, qualitatively in agreement with the findings in~\cite{Ayala:2015bgv} obtained in QCD at weak coupling. This phenomenon, the so-called inverse magnetic catalysis,
is shown in Fig.~\ref{invmagcat} where we show the decrease of (pseudo)critical chemical
potential ($\mu_c$) for chiral symmetry breaking with increasing temperature at a fixed magnetic field $\tl{e}B=0.5$ GeV$^2$.

\begin{figure}[!ht]
\subfloat[]{\includegraphics[scale=0.6]{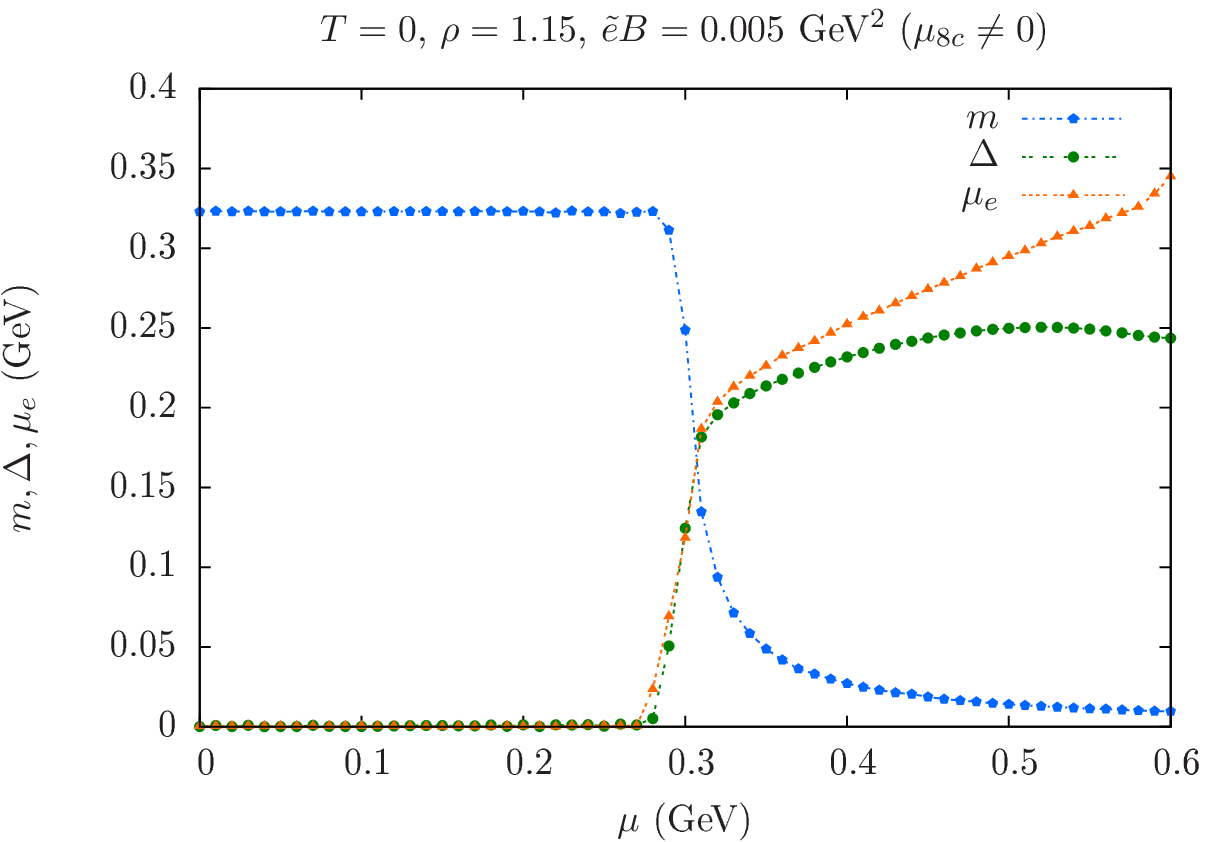}\label{f2a}}
\subfloat[]{\includegraphics[scale=0.6]{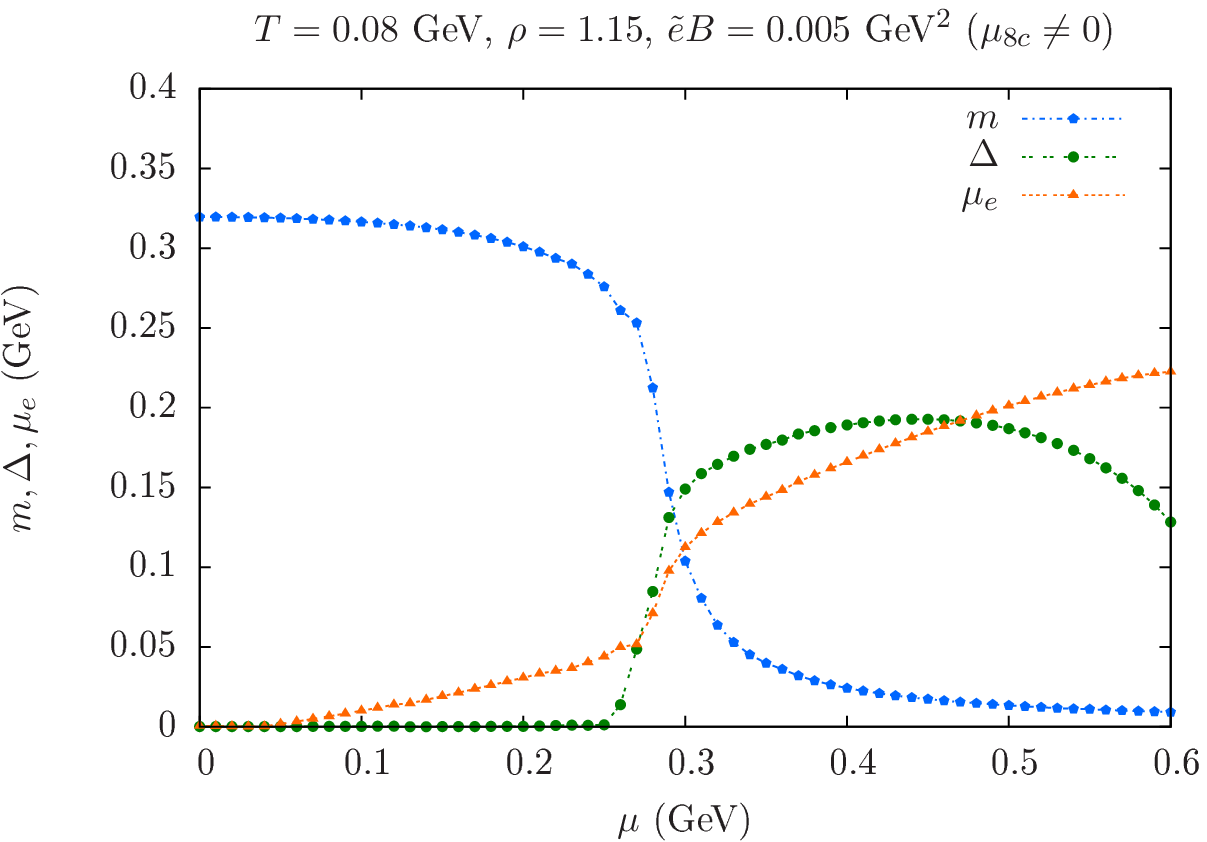}\label{f2b}}\\
\subfloat[]{\includegraphics[scale=0.6]{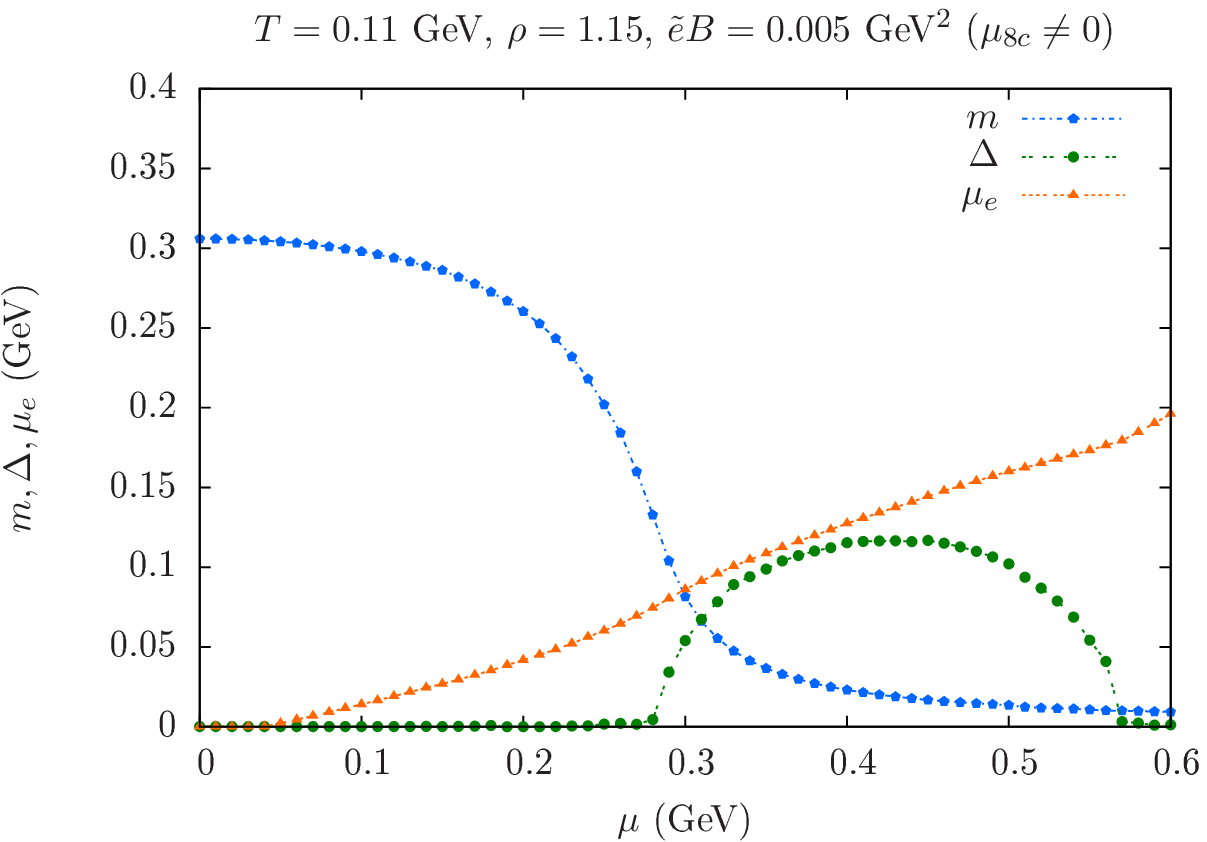}\label{f2c}}
\subfloat[]{\includegraphics[scale=0.6]{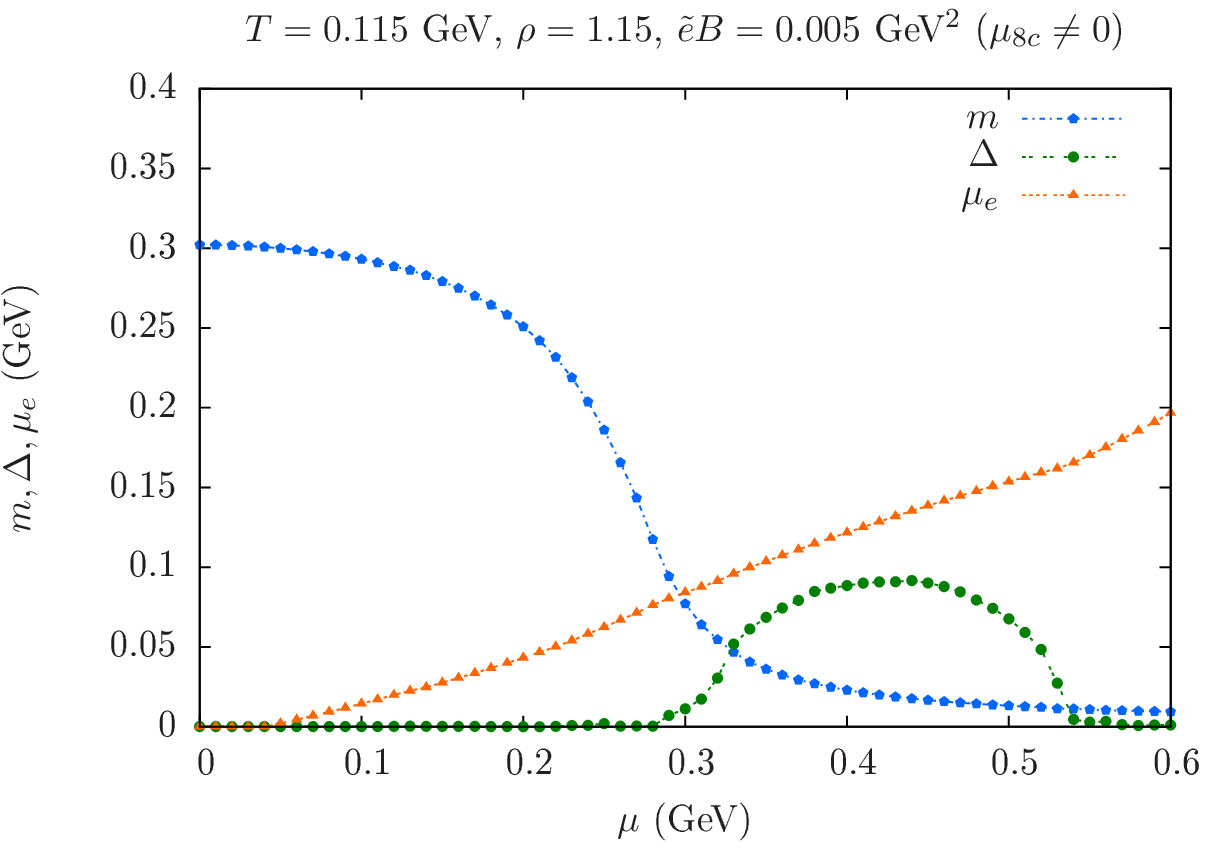}\label{f2d}}
\caption{The dependence of $m$, $\Delta$ and $\mu_e$ on $\mu$ for the
electric and color charge neutral quark matter in the weak magnetic field limit
for $\rho=1.15$. The panel shows four plots for $T=0,0.08,0.11$ and 0.115 GeV.}
\label{fig:2}
\end{figure}

Fig.~\ref{fig:2}'s panel extends the results in Fig.~\ref{fig:1} to the case
$\rho=1.15$, which at zero temperature and weak magnetic field yields the usual 2SC phase, i.e.,
$\Dl>\dl\mu$. The mismatch between the Fermi surfaces imposed by the neutrality condition at zero field is further increased by the application of a magnetic field, and this eventually makes the mismatch large enough to disrupt pairing (i.e., to satisfy the condition $\delta\mu > \Delta$). Because the gap is so large at strong coupling, the value of the applied field also needs to be large (approximately $\tl{e}B \gtrsim 0.4$ GeV$^2\approx 2\times 10^{19}$~G), which explains our earlier finding~\cite{Mandal:2012fq} that most quarks are in the lowest Landau level.
Such a large magnetic field
is at the upper limit allowed in compact stars as suggested by the virial theorem~\cite{Cardall:2000bs}, so it is unlikely that this effect is realized in their cores. An interesting feature in Fig.~\ref{fig:2}'s panel is 
that increasing temperature reduces the gap strongly while $\mu_e$ decreases only slightly in the gapped phase, so that the 2SC phase with $\Dl>\dl\mu$ 
is not favored even at large coupling. But again, this effect is only relevant for $T \gtrsim 0.1$ GeV at typical core densities, so that the CSC gap is likely to persist even in newly born compact stars that are hot.

We observe that the BCS gap $\Dl$ in the neutral phase is always smaller than the gap when charge neutrality is not enforced.
This is because the presence of $\mu_e$ in the neutral g2SC phase creates 
mismatch in the Fermi surfaces of two pairing quarks and suppresses pairing~\cite{Huang:2002zd,Huang:2003xd}. 
We can think of $\dl\mu=\mu_e/2$ as defined in Eq.~\eqref{eq:mismatch} 
as a Fermi surface mismatch parameter. It was shown in~\cite{Alford:2000ze}, for zero magnetic field and temperature, that if the mismatch $\dl\mu^0 > \Dl^0/\sqrt{2}$, a first order phase transition occurs and the superconducting gap vanishes. Beyond this point, a Larkin-Ovchinnikov-Fulde-Ferrell (LOFF) phase or another heterogeneous gapped phase may be favored. Here, $\dl\mu^0=\mu^0_e/2$ where $\mu^0_e$ is the electron chemical
potential in normal quark matter (i.e., $\Dl=0$) and $\Dl^0$ is the diquark gap without any charge neutrality. 
The condition $\dl\mu^0=\Dl^0/\sqrt{2}$ is similar to the famous Clogston-Chandrasekhar point in electronic superconductors~\cite{Clogston:1962zz,Chandrasekhar:1962zz}, except there is no analog of charge neutrality in condensed matter systems. 

\begin{figure}[!ht]
\subfloat[]{\includegraphics[scale=0.6]{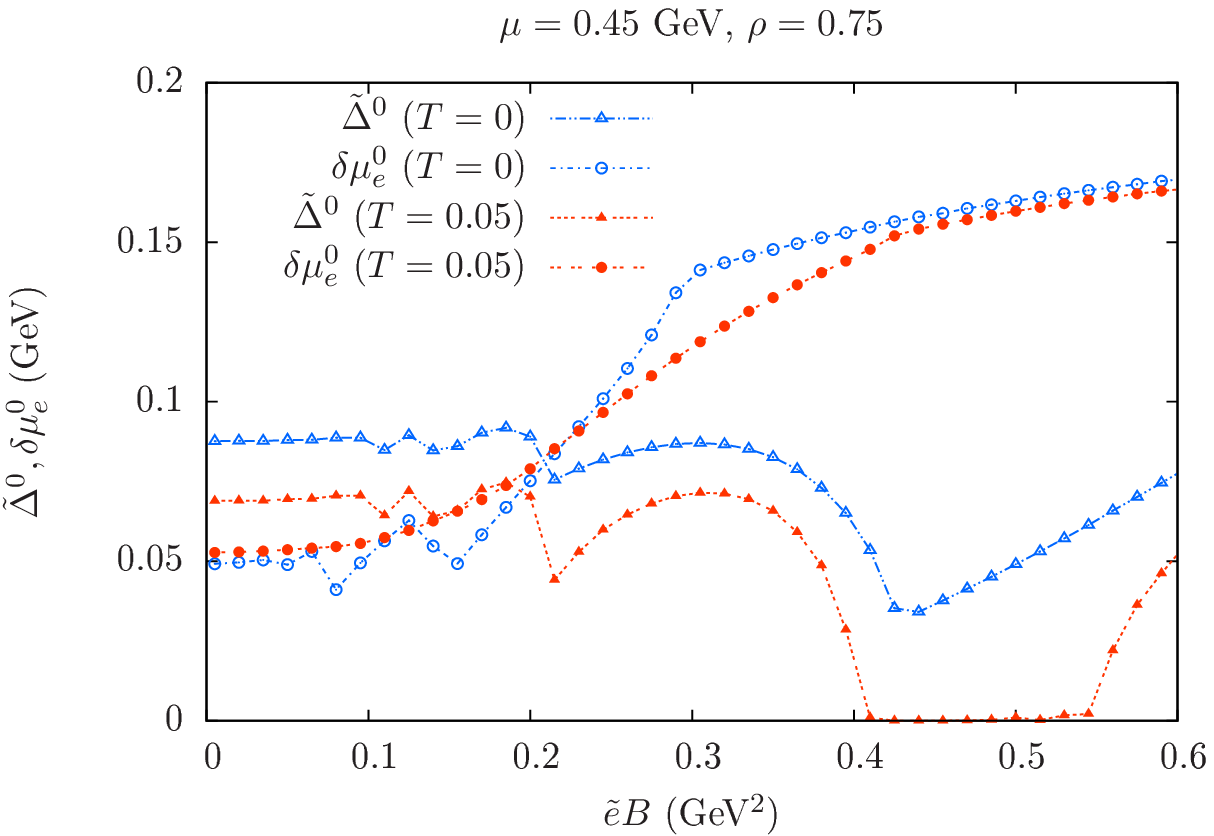}\label{f3a}}
\subfloat[]{\includegraphics[scale=0.6]{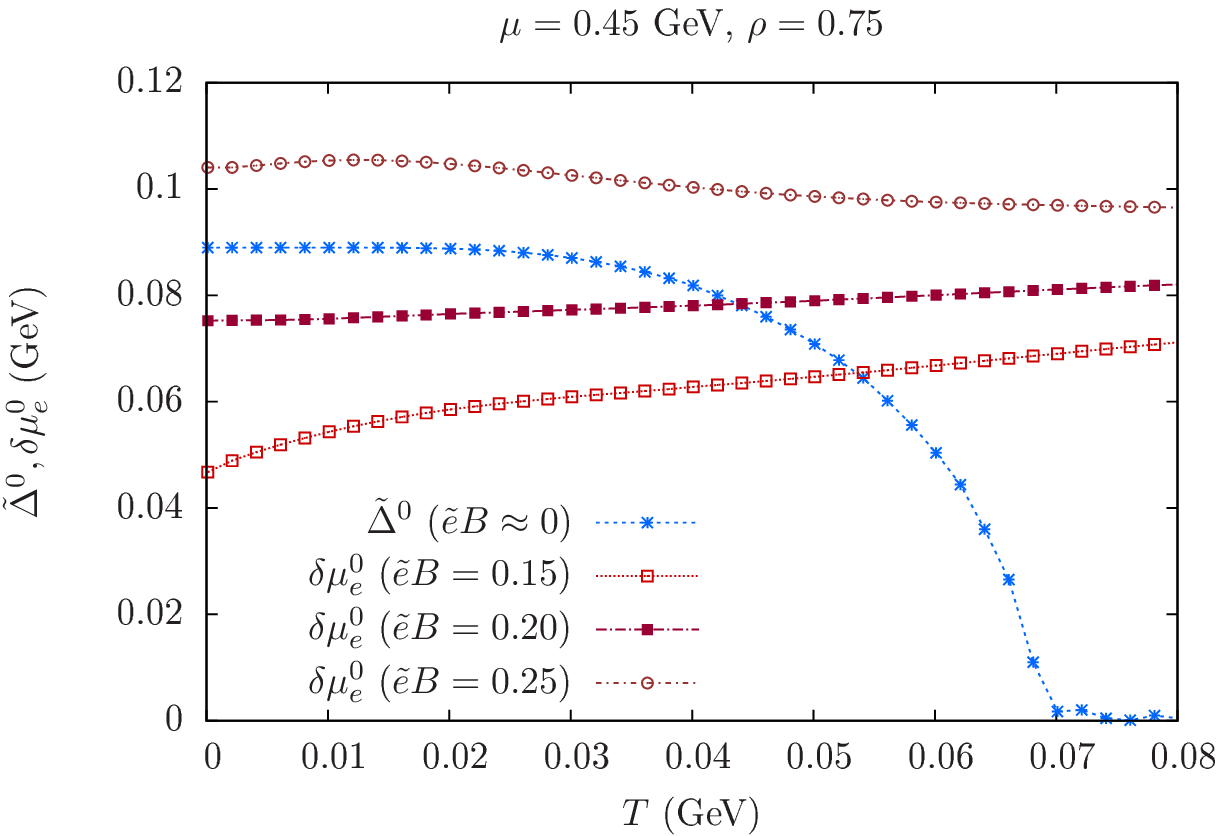}\label{f3b}}
\caption{Behavior of $\dl\mu^0_e=(\mu_e^0/2)$ and $\tl{\Dl}^0(=\Dl^0/\sqrt{2})$ as functions
of (a) $\tl{e}B$ and (b) $T$  at a fixed $\mu=0.45$ GeV for $\rho=0.75$.}
\label{f3}
\end{figure}

\begin{figure}[!ht]
\subfloat[]{\includegraphics[scale=0.6]{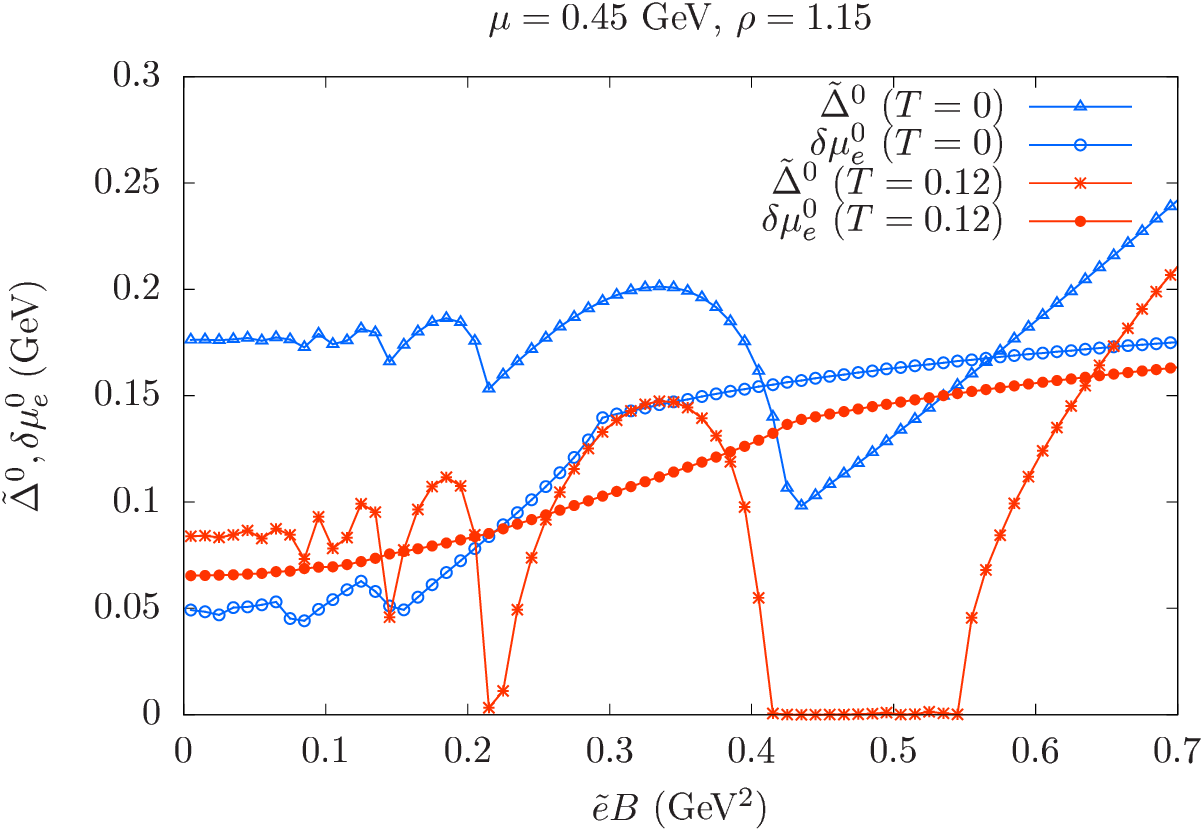}\label{f5a}}
\subfloat[]{\includegraphics[scale=0.6]{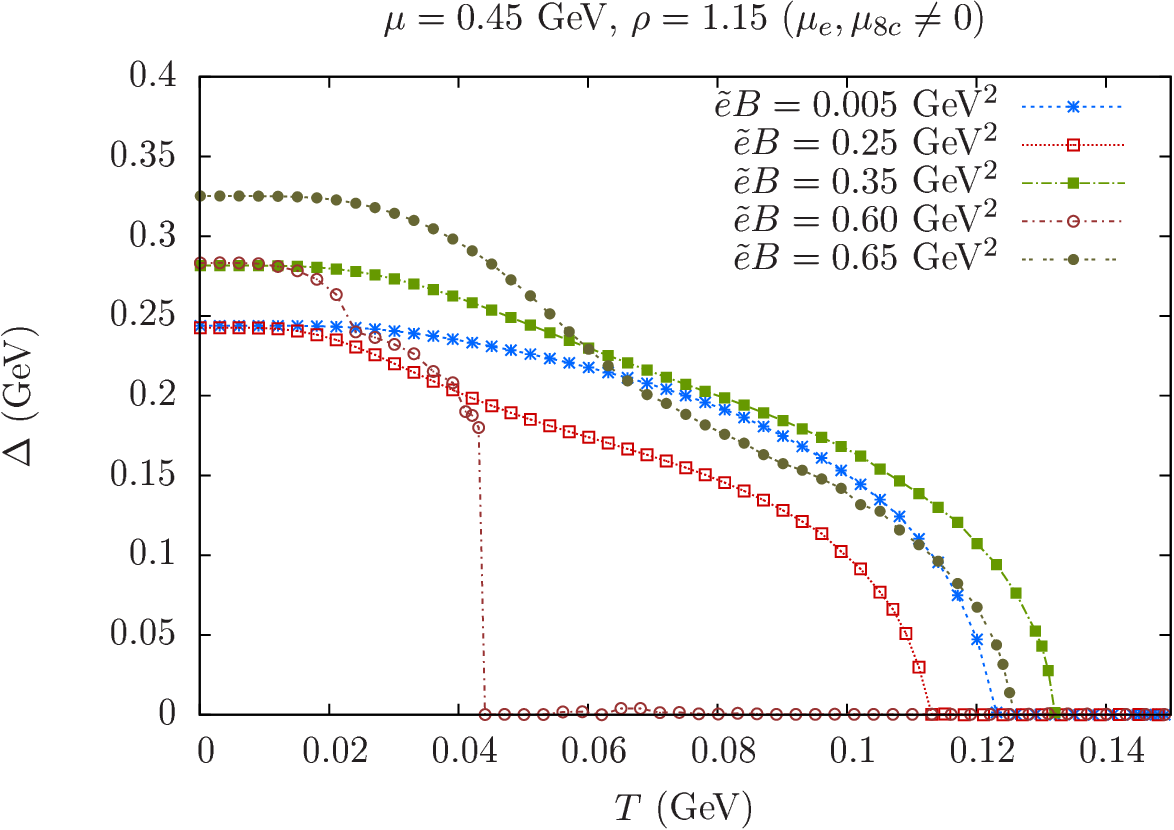}\label{f5b}}
\caption{Behavior of $\dl\mu^0_e=(\mu_e^0/2)$ and $\tl{\Dl}^0(=\Dl^0/\sqrt{2})$ as functions
of (a) $\tl{e}B$ and (b) $T$  at a fixed $\mu=0.45$ GeV for $\rho=1.15$.}
\label{f5}
\end{figure}

The results shown in Fig.~\ref{f3}'s panel extend this finding quantitatively into the regime of nonzero $B$ and $T$. 
In Fig.~\ref{f3a}, we compare $\dl\mu^0$ with $\tl{\Dl}^0$ (for notational ease, we define $\tl{\Dl}^0=\Dl^0/\sqrt{2}$) as functions
of $\tl{e}B$ for $T=0,0.05$ GeV. In Fig.~\ref{f3b}, we show 
similar comparison as functions of $T$ for $\tl{e}B=0.15,0.20,0.25$ GeV$^2$. 
We fix $\mu=0.45$ GeV and $\rho=0.75$ for both of these plots. We show $\tl{\Dl}^0$ only for $\tl{e}B\approx 0$ in Fig.~\ref{f3b} because $\tl{\Dl}^0$ 
is almost insensitive to the magnetic field as long as its magnitude is not too large, $\tl{e}B\lesssim 0.3$ GeV$^2$ (this can be seen later in Fig.~\ref{f5a}).
We plainly see that the condition $\dl\mu^0_e=\tl{\Dl}^0$, which marks
the breakdown of the homogeneous pairing ansatz, can happen both along
$\tl{e}B$ and $T$ directions. From Fig.~\ref{f3a}, we see that this happens 
approximately at a critical magnetic field $\tl{e}B_c\approx 0.21$ GeV$^2$ for $T=0$.
As we increase $T$, the critical value $\tl{e}B_c$ shifts to the lower side.
For instance, $\tl{e}B_c\approx 0.17$ GeV$^2$ for $T=0.05$ GeV.
The value of $\dl\mu^0$ is not very sensitive to $T$ (although oscillations along the
$\tl{e}B$ direction in $\dl\mu^0$ diminish with increasing $T$). However, $\tl{\Dl}^0$ decreases strongly with increasing 
$T$, and as a result, $\dl\mu^0$ and $\tl{\Dl}^0$ intersect at smaller values of 
$\tl{e}B$ with higher values of $T$. In fact, beyond a certain critical temperature
$T_c$, $\tl{\Dl}^0$ is always smaller than $\dl\mu^0$ for any $\tl{e}B$. Therefore,
for such $T>T_c$, the charge neutral gap $\Dl$ disappears. 

\begin{figure}[!ht]
\subfloat[]{\includegraphics[scale=0.6]{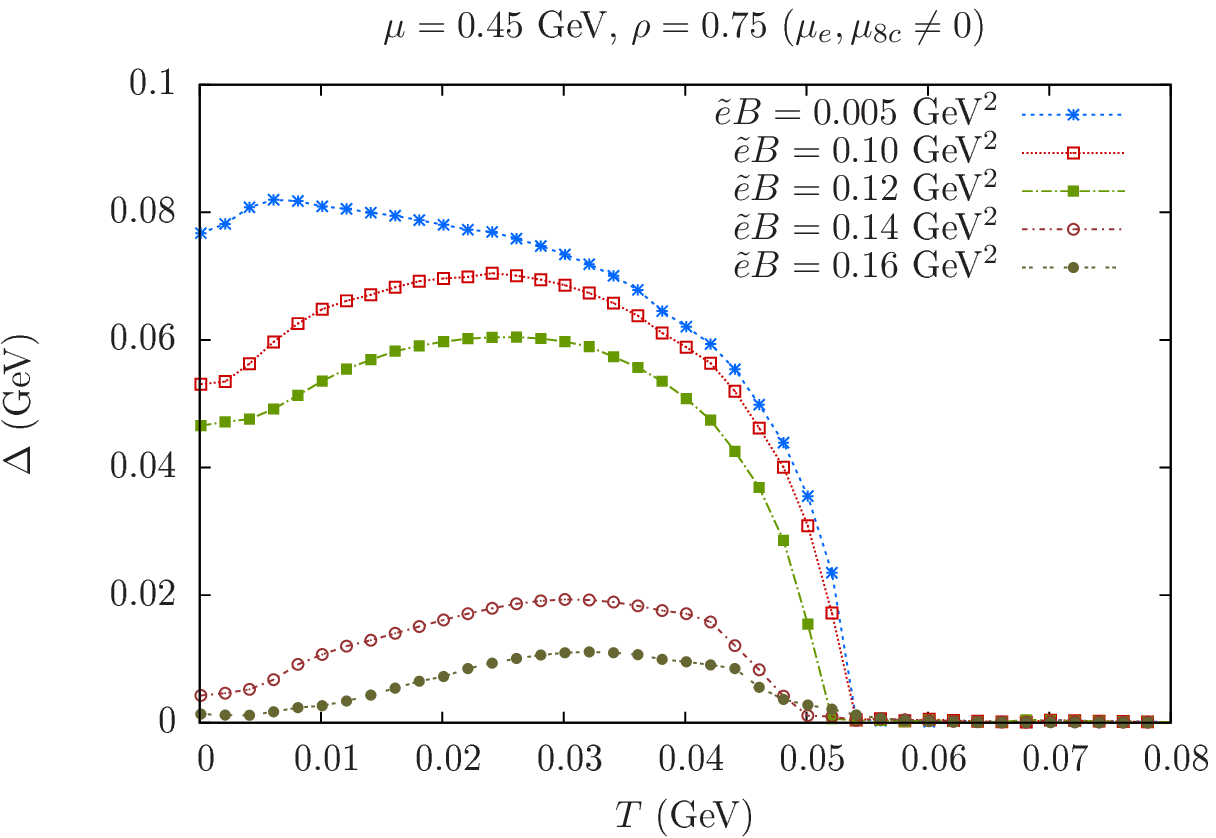}\label{f4a}}
\subfloat[]{\includegraphics[scale=0.6]{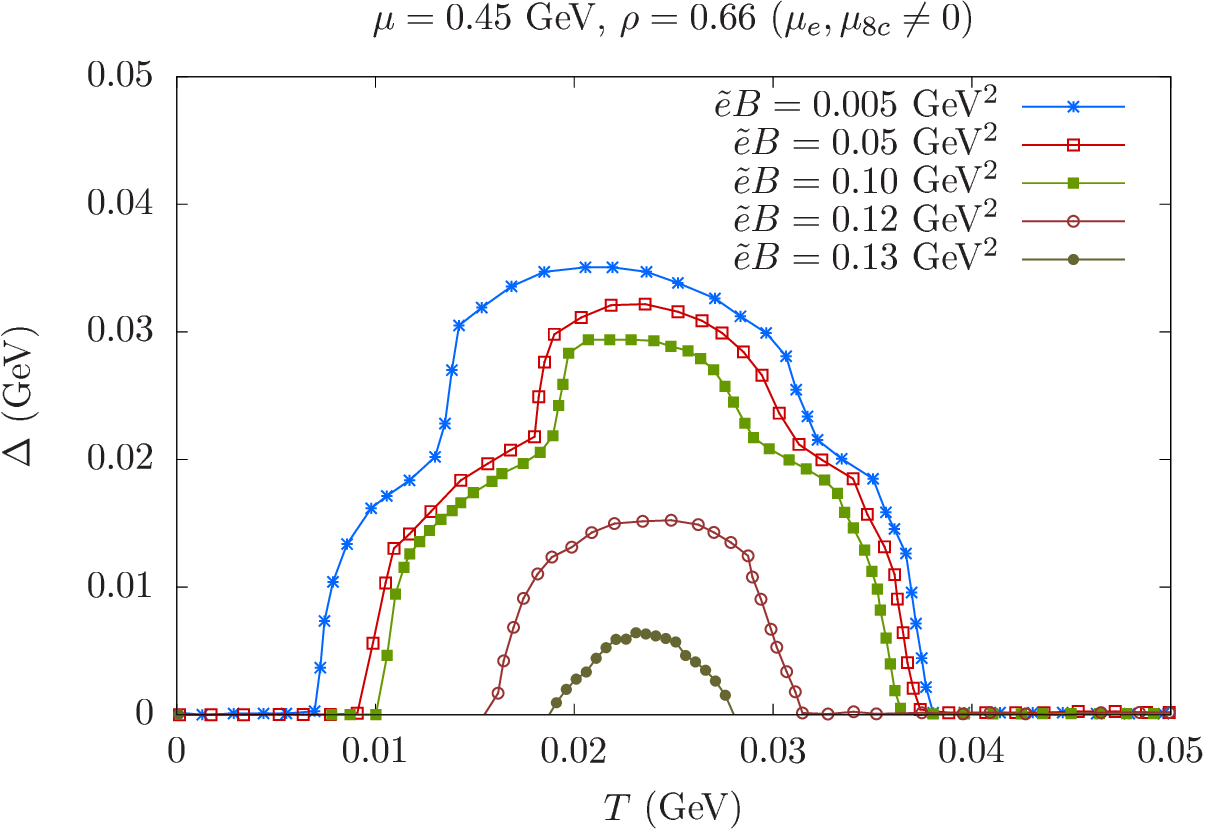}\label{f4b}}
\caption{Effect of the magnetic field on the temperature profile of the superconducting gap at $\mu=0.45$ GeV for (a) $\rho=0.75$ and (b) 
$\rho=0.66$.}
\label{f4}
\end{figure}

In Fig.~\ref{f3b}, we see how this 
$T_c$ depends on the magnetic field.
If we compare $\dl\mu^0(\tl{e}B\approx 0)$ with $\tl{\Dl}^0$ in Fig.~\ref{f3b},
we see they intersect at a temperature around $T_c\approx 0.054$ GeV. 
As we increase $\tl{e}B$, the critical value $T_c$ shifts to the lower side.
For instance, $T_c\approx 0.043$ GeV for $\tl{e}B=0.2$ GeV$^2$. The diquark gap $\Dl^0$ is not very sensitive to $\tl{e}B$, but $\mu_e^0$ (which equals $2\delta\mu^0$) increases with increasing 
$\tl{e}B$. This is because the difference in the number densities of the pairing quarks $(n_{d,g}-n_{u,r})$ or 
$(n_{d,r}-n_{u,g})$ is proportional in $\tl{e}B$ because 
of phase space crowding. As a result of the positive correlation between $\tl{e}B$ and the mismatch, $\dl\mu^0$ and $\tl{\Dl}^0$ intersect at smaller values of 
$T$ as $\tl{e}B$ increases,  and there exists a certain critical magnetic field $\tl{e}B_c$ beyond which 
no pairing is possible (this can also be seen in Fig.~\ref{f4a}). The main message from Figs.~\ref{f3a} and 
\ref{f3b} is that the g2SC pairing is disturbed with both increasing 
$\tl{e}B$ and $T$, but for different reasons, the former due to Fermi surface mismatch and the latter due to thermal effects.
One final interesting observation in Fig.~\ref{f3a} is that for $T=0.05$ GeV, there exist two critical magnetic fields,
$\tl{e}B_{c_1}\approx 0.41$ GeV$^2$ for the transition from CSC to normal
phase and $\tl{e}B_{c_2}\approx 0.54$ GeV$^2$ for the transition from normal
to CSC phase. This behavior is already observed in Ref.~\cite{Fayazbakhsh:2010bh}
and this happens due to large oscillations in the diquark gap. It is worth mentioning that such oscillations may be amplified upon inclusion of gluonic effects~\cite{Sinha:2013dfa}, which are lacking in our model.

In Figs.~\ref{f4a} and \ref{f4b}, we show $\Dl$
(with charge neutrality) as functions of $T$ for different magnetic fields for $\rho=0.75$ and $\rho=0.66$, respectively. 
For these plots, we fix $\mu=0.45$ GeV. The intercepts on the axes of Fig.~\ref{f4a} reveal the nonuniversal 
behavior of the ratio $T_c/\Dl^0$ in the neutral g2SC phase, which was noted already in~\cite{Huang:2003xd} 
but without the magnetic field.
We have checked that the universal relation $T_c/\Dl^0\approx 0.57$ is restored when neutrality is not enforced, with no critical magnetic field disrupting the pairing, confirming the findings in Ref.~\cite{Fayazbakhsh:2010bh}. 
Fig.~\ref{f4b} shows that the finite-temperature CSC gap is nonvanishing even if the zero-temperature gap value vanishes, since thermal effects help to bridge the mismatch in the $u$ and $d$ Fermi surfaces. These results extend the findings of~\cite{Huang:2003xd} to the case of strong magnetic field, which increases the stress on pairing, lowering the finite-temperature gap. This is true for other values of $\rho$ as well.

Figs.~\ref{f5a} and \ref{f5b} are similar plots to Figs.~\ref{f3a} and \ref{f3b} but for $\rho=1.15$.
In our previous work in Ref.~\cite{Mandal:2012fq}, we showed that for large $\rho$, 
$\tl{\Dl}^0$ can become smaller than $\dl\mu^0$ in some window of $\tl{e}B$ due to large 
oscillation in the superconducting gap. In Fig.~\ref{f5a} this happens in the range
$\tl{e}B\sim 0.41-0.57$ at $T=0$. As explained earlier in this section, the diquark gap disappears in those
ranges of $\tl{e}B$. This region becomes wider as we increase $T$, {\it e.g.},
$\tl{e}B\sim 0.38-0.63$ for $T=0.12$ GeV. A direct consequence of this is shown in Fig.~\ref{f5b}, where $\Dl$ abruptly vanishes at $T\sim 0.04$ GeV 
for a particular choice, $\tl{e}B \sim 0.6$ GeV$^2$. As we increase T, the
upper limiting value for the $\tl{e}B$ region where $\tl{\Dl}^0< \dl\mu^0_e$ expands,
and happens to reach the value $\tl{e}B = 0.6$ GeV$^2$ for $T=0.04$ GeV. 
However, the strength of the magnetic field is again quite extreme for this effect to be realized in compact stars. 
Thus, we expect that while the magnitude of the charge neutral gap in the quark cores of strongly 
magnetized neutron stars can be strongly dependent on the temperature, the magnetic field is probably not large enough to
show any dramatic effect, especially a complete suppression of the gap. Recently, it was pointed out that the magnetic field could be large enough in magnetars to suppress ordinary nucleonic superconductivity~\cite{Sinha:2015bva}. Of course, it is still likely that the magnetic field will lead to significant effects on the cooling of quark matter through anisotropy-modified transport properties~\cite{Huang:2009ue}
and the spectrum of quasiparticles participating in the cooling processes~\cite{Anand:1979uq}.


\section{Conclusions}
\label{sec:conc}

In this paper, we have studied the effect of temperature (in the tens of MeV range) on 
charge neutral 2SC quark matter subject to a constant external magnetic field. The NJL 
model, being well adapted to superconductivity, is used to numerically determine the 
chiral and the diquark condensates from the gap equations under neutrality conditions. Our 
results extend prior studies of the competition between chiral symmetry breaking and color 
superconductivity to the regime of large magnetic field and high temperature expected in the
core of newly born compact stars.  Our results in the limit of either zero temperature or zero magnetic field 
confirm the findings of~\cite{Huang:2002zd,Fayazbakhsh:2010gc,Mandal:2012fq,Mandal:2009uk,Huang:2003xd}. 
However, we find an interesting interplay of these two parameters ($T$ and $B$) in a 
charge neutral color superconductor that is new and constitutes the main finding of 
this paper, summarized below.

The g2SC phase in magnetized hot quark matter is characterized by a magnetic field-dependent critical pairing temperature $T_c$ beyond which the charge neutral gap disappears. This is due to the strong negative effect of the magnetic field on pairing of mismatched Fermi surfaces resulting from the charge neutrality condition.
The finite-temperature CSC gap in charge neutral two-flavor quark matter can be nonzero and large even if the zero-temperature gap vanishes or is very small. First noted in the context of nonsuperconducting matter (weak coupling) in~\cite{Huang:2003xd}, our results show that this anomalous behavior can also manifest in the g2SC phase (moderate coupling) at large magnetic field. In the 2SC phase (strong coupling), we do not expect or find such behavior, but we do observe that the gap can vanish abruptly beyond a certain (fixed) temperature for a critical magnetic field, and reappear at higher magnetic field. This indicates that the homogeneous gap in the charge neutral case is severely stressed by both increasing $T$ and $B$, and that more complicated order parameters, for example the LOFF state~\cite{Fulde:1964zz,Larkin:1964zz}, could be more robust alternatives to 2SC pairing in the quark cores of newly born magnetars. 
 
We also find that the ``mixed broken phase'', which refers to the region of coexisting chiral and diquark condensates is more sensitive to the magnetic field value and temperature than the imposition of neutrality conditions. Strictly speaking, chiral symmetry is never restored due to the explicit quark mass, so the terminology of the mixed phase used in~\cite{Huang:2002zd} is somewhat misleading. However, Fig.~\ref{fig:2} suggests there is indeed a window of density where both condensates have nonzero VEVs, in agreement with~\cite{Huang:2002zd}, which did not impose neutrality. Increasing the magnetic field shrinks this window, as does decreasing the temperature. Therefore, we conclude that the mixed broken phase is stabilized by the opposing effect of temperature and magnetic field, and charge imbalance due to neutrality does not destroy the mixed phase, as claimed 
in~\cite{Mandal:2012fq}. This has interesting consequences for the interior of compact stars: if the phase transition to hot quark matter in the core of a magnetized star creates first a metastable state which
then drops to the true minimum (color superconductor), nucleation of (partially) chirally restored and superconducting droplets can
happen simultaneously. In addition to the color-magnetic flux tubes expected to thread two-flavor superconductors at the femtoscale already at much lower field values~\cite{Alford:2010qf}, a magnetic field of order $\tilde{e}B/\mu^2\sim 1$ is likely to show some local variation on much larger scales in the initial stages of the formation of the dense compact star due to fluid turbulence and rapid rotation. In this case, magnetic domains with different magnetization can form. However, this is not a new idea, since it is already known that in the presence of a strong
magnetic field the color superconducting Cooper pairs acquire magnetic
moments \cite{Feng:2011fj}, whose polarizations could, in principle, be the source of such magnetic domains. In any case, such kinds of nucleation and domain formation processes release latent heat that is very
large owing to the large value of the magnetic field, and constitute a potential mechanism for gamma-ray bursts of protoquark star~\cite{Ouyed:2005dz} or protomagnetar origin~\cite{Thompson:2006ii}.

\section*{ACKNOWLEDGMENTS}

T.M. is supported by funding from the Carl Trygger Foundation under Contract No. CTS-14:206 and 
the Swedish Research Council under Contract No. 621-2011-5107.

\bibliography{cnmag2SC}{}

\begin{thebibliography}{10}

\bibitem{Barrois:1977xd}
Bertrand~C. Barrois.
\newblock {Superconducting Quark Matter}.
\newblock {\em Nucl. Phys.}, B129:390--396, 1977.

\bibitem{Bailin:1983bm}
D.~Bailin and A.~Love.
\newblock {Superfluidity and Superconductivity in Relativistic Fermion
  Systems}.
\newblock {\em Phys. Rept.}, 107:325, 1984.

\bibitem{Iwasaki:1994ij}
M.~Iwasaki and T.~Iwado.
\newblock {Superconductivity in the quark matter}.
\newblock {\em Phys. Lett.}, B350:163--168, 1995.

\bibitem{Collins:1974ky}
John~C. Collins and M.~J. Perry.
\newblock {Superdense Matter: Neutrons Or Asymptotically Free Quarks?}
\newblock {\em Phys. Rev. Lett.}, 34:1353, 1975.

\bibitem{Alford:1997zt}
Mark~G. Alford, Krishna Rajagopal, and Frank Wilczek.
\newblock {QCD at finite baryon density: Nucleon droplets and color
  superconductivity}.
\newblock {\em Phys. Lett.}, B422:247--256, 1998.

\bibitem{Rapp:1997zu}
R.~Rapp, Thomas Schäfer, Edward~V. Shuryak, and M.~Velkovsky.
\newblock {Diquark Bose condensates in high density matter and instantons}.
\newblock {\em Phys. Rev. Lett.}, 81:53--56, 1998.

\bibitem{Berges:1998rc}
Juergen Berges and Krishna Rajagopal.
\newblock {Color superconductivity and chiral symmetry restoration at nonzero
  baryon density and temperature}.
\newblock {\em Nucl. Phys.}, B538:215--232, 1999.

\bibitem{Alford:1998mk}
Mark~G. Alford, Krishna Rajagopal, and Frank Wilczek.
\newblock {Color flavor locking and chiral symmetry breaking in high density
  QCD}.
\newblock {\em Nucl. Phys.}, B537:443--458, 1999.

\bibitem{Son:1998uk}
D.~T. Son.
\newblock {Superconductivity by long range color magnetic interaction in high
  density quark matter}.
\newblock {\em Phys. Rev.}, D59:094019, 1999.

\bibitem{Alford:2007xm}
Mark~G. Alford, Andreas Schmitt, Krishna Rajagopal, and Thomas Schäfer.
\newblock {Color superconductivity in dense quark matter}.
\newblock {\em Rev. Mod. Phys.}, 80:1455--1515, 2008.

\bibitem{Manuel:2007pz}
Cristina Manuel and Felipe~J. Llanes-Estrada.
\newblock {Bulk viscosity in a cold CFL superfluid}.
\newblock {\em JCAP}, 0708:001, 2007.

\bibitem{Alford:2008pb}
Mark~G. Alford, Matt Braby, and Andreas Schmitt.
\newblock {Bulk viscosity in kaon-condensed color-flavor locked quark matter}.
\newblock {\em J. Phys.}, G35:115007, 2008.

\bibitem{Brauner:2009df}
Tomas Brauner, Jin-yi Pang, and Qun Wang.
\newblock {Symmetry breaking patterns and collective modes of spin-one color
  superconductors}.
\newblock {\em Nucl. Phys.}, A844:216C--223C, 2010.

\bibitem{Braby:2009dw}
Matt Braby, Jingyi Chao, and Thomas Schäfer.
\newblock {Thermal conductivity of color-flavor locked quark matter}.
\newblock {\em Phys. Rev.}, C81:045205, 2010.

\bibitem{Alford:2014doa}
Mark~G. Alford, Hiromichi Nishimura, and Armen Sedrakian.
\newblock {Transport coefficients of two-flavor superconducting quark matter}.
\newblock {\em Phys. Rev.}, C90(5):055205, 2014.

\bibitem{Weissenborn:2011qu}
Simon Weissenborn, Irina Sagert, Giuseppe Pagliara, Matthias Hempel, and Jurgen
  Schaffner-Bielich.
\newblock {Quark Matter In Massive Neutron Stars}.
\newblock {\em Astrophys. J.}, 740:L14, 2011.

\bibitem{Klahn:2013kga}
T.~Klähn, R.~Łastowiecki, and D.~B. Blaschke.
\newblock {Implications of the measurement of pulsars with two solar masses for
  quark matter in compact stars and heavy-ion collisions: A
  Nambu–Jona-Lasinio model case study}.
\newblock {\em Phys. Rev.}, D88(8):085001, 2013.

\bibitem{Ferrer:2015vca}
E.~J. Ferrer, V.~de~la Incera, and L.~Paulucci.
\newblock {Gluon effects on the equation of state of color superconducting
  strange stars}.
\newblock {\em Phys. Rev.}, D92(4):043010, 2015.

\bibitem{Duncan:1992hi}
Robert~C. Duncan and Christopher Thompson.
\newblock {Formation of very strongly magnetized neutron stars - implications
  for gamma-ray bursts}.
\newblock {\em Astrophys. J.}, 392:L9, 1992.

\bibitem{Paczynski:1992zz}
Bohdan Paczynski.
\newblock {GB 790305 as a very strongly magnetized neutron star}.
\newblock {\em Acta Astron.}, 42:145--153, 1992.

\bibitem{Guver:2007ky}
Tolga Guver, Feryal Ozel, Ersin Gogus, and Chryssa Kouveliotou.
\newblock {The Magnetar Nature and the Outburst Mechanism of a Transient
  Anomalous X-ray Pulsar}.
\newblock {\em Astrophys. J.}, 667:L73, 2007.

\bibitem{Ferrer:2010wz}
Efrain~J. Ferrer, Vivian de~la Incera, Jason~P. Keith, Israel Portillo, and
  Paul~L. Springsteen.
\newblock {Equation of State of a Dense and Magnetized Fermion System}.
\newblock {\em Phys. Rev.}, C82:065802, 2010.

\bibitem{Cardall:2000bs}
Christian~Y. Cardall, Madappa Prakash, and James~M. Lattimer.
\newblock {Effects of strong magnetic fields on neutron star structure}.
\newblock {\em Astrophys. J.}, 554:322--339, 2001.

\bibitem{Ferrer:2005vd}
Efrain~J. Ferrer, Vivian de~la Incera, and Cristina Manuel.
\newblock {Magnetic color flavor locking phase in high density QCD}.
\newblock {\em Phys. Rev. Lett.}, 95:152002, 2005.

\bibitem{Ferrer:2006vw}
Efrain~J. Ferrer, Vivian de~la Incera, and Cristina Manuel.
\newblock {Color-superconducting gap in the presence of a magnetic field}.
\newblock {\em Nucl. Phys.}, B747:88--112, 2006.

\bibitem{Ferrer:2007iw}
Efrain~J. Ferrer and Vivian de~la Incera.
\newblock {Magnetic Phases in Three-Flavor Color Superconductivity}.
\newblock {\em Phys. Rev.}, D76:045011, 2007.

\bibitem{Fukushima:2007fc}
Kenji Fukushima and Harmen~J. Warringa.
\newblock {Color superconducting matter in a magnetic field}.
\newblock {\em Phys. Rev. Lett.}, 100:032007, 2008.

\bibitem{Noronha:2007wg}
Jorge~L. Noronha and Igor~A. Shovkovy.
\newblock {Color-flavor locked superconductor in a magnetic field}.
\newblock {\em Phys. Rev.}, D76:105030, 2007.
\newblock [Erratum: Phys. Rev.D86,049901(2012)].

\bibitem{Ferrer:2013noa}
Efrain~J. Ferrer, Vivian de~la Incera, Israel Portillo, and Matthew Quiroz.
\newblock {New look at the QCD ground state in a magnetic field}.
\newblock {\em Phys. Rev.}, D89(8):085034, 2014.

\bibitem{Kharzeev:2007jp}
Dmitri~E. Kharzeev, Larry~D. McLerran, and Harmen~J. Warringa.
\newblock {The Effects of topological charge change in heavy ion collisions:
  'Event by event P and CP violation'}.
\newblock {\em Nucl. Phys.}, A803:227--253, 2008.

\bibitem{Skokov:2009qp}
V.~Skokov, A.~{\relax Yu}. Illarionov, and V.~Toneev.
\newblock {Estimate of the magnetic field strength in heavy-ion collisions}.
\newblock {\em Int. J. Mod. Phys.}, A24:5925--5932, 2009.

\bibitem{Voronyuk:2011jd}
V.~Voronyuk, V.~D. Toneev, W.~Cassing, E.~L. Bratkovskaya, V.~P. Konchakovski,
  and S.~A. Voloshin.
\newblock {(Electro-)Magnetic field evolution in relativistic heavy-ion
  collisions}.
\newblock {\em Phys. Rev.}, C83:054911, 2011.

\bibitem{Gusynin:1994re}
V.~P. Gusynin, V.~A. Miransky, and I.~A. Shovkovy.
\newblock {Catalysis of dynamical flavor symmetry breaking by a magnetic field
  in (2+1)-dimensions}.
\newblock {\em Phys. Rev. Lett.}, 73:3499--3502, 1994.
\newblock [Erratum: Phys. Rev. Lett.76,1005(1996)].

\bibitem{Semenoff:1998bk}
G.~W. Semenoff, I.~A. Shovkovy, and L.~C.~R. Wijewardhana.
\newblock {Phase transition induced by a magnetic field}.
\newblock {\em Mod. Phys. Lett.}, A13:1143--1154, 1998.

\bibitem{Ferrer:1999gs}
E.~J. Ferrer, V.~P. Gusynin, and V.~de~la Incera.
\newblock {Boundary effects in the magnetic catalysis of chiral symmetry
  breaking}.
\newblock {\em Phys. Lett.}, B455:217--223, 1999.

\bibitem{Kabat:2002er}
Daniel~N. Kabat, Ki-Myeong Lee, and Erick~J. Weinberg.
\newblock {QCD vacuum structure in strong magnetic fields}.
\newblock {\em Phys. Rev.}, D66:014004, 2002.

\bibitem{Miransky:2002rp}
V.~A. Miransky and I.~A. Shovkovy.
\newblock {Magnetic catalysis and anisotropic confinement in QCD}.
\newblock {\em Phys. Rev.}, D66:045006, 2002.

\bibitem{Shovkovy:2012zn}
Igor~A. Shovkovy.
\newblock {Magnetic Catalysis: A Review}.
\newblock {\em Lect. Notes Phys.}, 871:13--49, 2013.

\bibitem{Avancini:2011zz}
S.~S. Avancini, D.~P. Menezes, and C.~Providencia.
\newblock {Finite temperature quark matter under strong magnetic fields}.
\newblock {\em Phys. Rev.}, C83:065805, 2011.

\bibitem{Fukushima:2012kc}
Kenji Fukushima and Yoshimasa Hidaka.
\newblock {Magnetic Catalysis Versus Magnetic Inhibition}.
\newblock {\em Phys. Rev. Lett.}, 110(3):031601, 2013.

\bibitem{Bali:2013esa}
G.~S. Bali, F.~Bruckmann, G.~Endrodi, F.~Gruber, and A.~Schaefer.
\newblock {Magnetic field-induced gluonic (inverse) catalysis and pressure
  (an)isotropy in QCD}.
\newblock {\em JHEP}, 04:130, 2013.

\bibitem{Mueller:2015fka}
Niklas Mueller and Jan~M. Pawlowski.
\newblock {Magnetic catalysis and inverse magnetic catalysis in QCD}.
\newblock {\em Phys. Rev.}, D91(11):116010, 2015.

\bibitem{Ahmad:2016iez}
Aftab Ahmad and Alfredo Raya.
\newblock {Inverse magnetic catalysis and confinement within a contact
  interaction model for quarks}.
\newblock {\em J. Phys.}, G43(6):065002, 2016.

\bibitem{Gorbar:2009bm}
E.~V. Gorbar, V.~A. Miransky, and I.~A. Shovkovy.
\newblock {Chiral asymmetry of the Fermi surface in dense relativistic matter
  in a magnetic field}.
\newblock {\em Phys. Rev.}, C80:032801, 2009.

\bibitem{Preis:2010cq}
Florian Preis, Anton Rebhan, and Andreas Schmitt.
\newblock {Inverse magnetic catalysis in dense holographic matter}.
\newblock {\em JHEP}, 03:033, 2011.

\bibitem{Chatterjee:2011yi}
Bhaswar Chatterjee, Hiranmaya Mishra, and Amruta Mishra.
\newblock {Chiral symmety breaking in 3-flavor Nambu-Jona Lasinio model in
  magnetic background}.
\newblock {\em Nucl. Phys.}, A862-863:312--315, 2011.
\newblock arXiv:1102.0875 [hep-ph].

\bibitem{Kharzeev:2013jha}
Dmitri Kharzeev, Karl Landsteiner, Andreas Schmitt, and Ho-Ung Yee.
\newblock {Strongly Interacting Matter in Magnetic Fields}.
\newblock {\em Lect. Notes Phys.}, 871:pp.1--624, 2013.

\bibitem{Mishra:2004gw}
Amruta Mishra and Hiranmaya Mishra.
\newblock {Color superconducting 2SC+s quark matter and gapless modes at finite
  temperatures}.
\newblock {\em Phys. Rev.}, D71:074023, 2005.

\bibitem{Ruester:2005jc}
Stefan~B. Ruester, Verena Werth, Michael Buballa, Igor~A. Shovkovy, and Dirk~H.
  Rischke.
\newblock {The Phase diagram of neutral quark matter: Self-consistent treatment
  of quark masses}.
\newblock {\em Phys. Rev.}, D72:034004, 2005.

\bibitem{Blaschke:2005uj}
D.~Blaschke, S.~Fredriksson, H.~Grigorian, A.~M. Oztas, and F.~Sandin.
\newblock {The Phase diagram of three-flavor quark matter under compact star
  constraints}.
\newblock {\em Phys. Rev.}, D72:065020, 2005.

\bibitem{Mandal:2012fq}
Tanumoy Mandal and Prashanth Jaikumar.
\newblock {Neutrality of a magnetized two-flavor quark superconductor}.
\newblock {\em Phys. Rev.}, C87:045208, 2013.

\bibitem{Huang:2003xd}
Mei Huang and Igor Shovkovy.
\newblock {Gapless color superconductivity at zero and at finite temperature}.
\newblock {\em Nucl. Phys.}, A729:835--863, 2003.

\bibitem{Fayazbakhsh:2010bh}
Sh. Fayazbakhsh and N.~Sadooghi.
\newblock {Phase diagram of hot magnetized two-flavor color superconducting
  quark matter}.
\newblock {\em Phys. Rev.}, D83:025026, 2011.

\bibitem{Andersen:2007qv}
Jens~O. Andersen and Lars Kyllingstad.
\newblock {Pion Condensation in a two-flavor NJL model: the role of charge
  neutrality}.
\newblock {\em J. Phys.}, G37:015003, 2009.

\bibitem{Huang:2001yw}
Mei Huang, Peng-fei Zhuang, and Wei-qin Chao.
\newblock {Massive quark propagator and competition between chiral and diquark
  condensate}.
\newblock {\em Phys. Rev.}, D65:076012, 2002.

\bibitem{Fayazbakhsh:2010gc}
Sh. Fayazbakhsh and N.~Sadooghi.
\newblock {Color neutral 2SC phase of cold and dense quark matter in the
  presence of constant magnetic fields}.
\newblock {\em Phys. Rev.}, D82:045010, 2010.

\bibitem{Frasca:2011zn}
Marco Frasca and Marco Ruggieri.
\newblock {Magnetic Susceptibility of the Quark Condensate and Polarization
  from Chiral Models}.
\newblock {\em Phys. Rev.}, D83:094024, 2011.

\bibitem{Allen:2015paa}
Pablo~G. Allen, Ana~G. Grunfeld, and Norberto~N. Scoccola.
\newblock {Magnetized color superconducting cold quark matter within the
  SU(2)$_f$ NJL model: A novel regularization scheme}.
\newblock {\em Phys. Rev.}, D92(7):074041, 2015.

\bibitem{Huang:2002zd}
Mei Huang, Peng-fei Zhuang, and Wei-qin Chao.
\newblock {Charge neutrality effects on 2 flavor color superconductivity}.
\newblock {\em Phys. Rev.}, D67:065015, 2003.

\bibitem{Ebert:1991pz}
D.~Ebert, L.~Kaschluhn, and G.~Kastelewicz.
\newblock {Effective meson - diquark Lagrangian and mass formulas from the
  Nambu-Jona-Lasinio model}.
\newblock {\em Phys. Lett.}, B264:420--425, 1991.

\bibitem{Mandal:2009uk}
Tanumoy Mandal, Prashanth Jaikumar, and Sanatan Digal.
\newblock {Chiral and Diquark condensates at large magnetic field in two-flavor
  superconducting quark matter}.
\newblock {arXiv:0912.1413 [nucl-th]}.

\bibitem{Ayala:2015bgv}
Alejandro Ayala, C.~A. Dominguez, L.~A. Hernandez, M.~Loewe, and R.~Zamora.
\newblock {Inverse magnetic catalysis from the properties of the QCD coupling
  in a magnetic field}.
\newblock {\em Phys. Lett.}, B759:99--103, 2016.

\bibitem{Alford:2000ze}
Mark~G. Alford, Jeffrey~A. Bowers, and Krishna Rajagopal.
\newblock {Crystalline color superconductivity}.
\newblock {\em Phys. Rev.}, D63:074016, 2001.

\bibitem{Clogston:1962zz}
A.~M. Clogston.
\newblock {Upper Limit for the Critical Field in Hard Superconductors}.
\newblock {\em Phys. Rev. Lett.}, 9:266--267, 1962.

\bibitem{Chandrasekhar:1962zz}
B.~S. Chandrasekhar.
\newblock A note on the maximum critical field of high‐field superconductors.
\newblock {\em Applied Physics Letters}, 1(1):7--8, 1962.

\bibitem{Sinha:2013dfa}
Monika Sinha, Xu-Guang Huang, and Armen Sedrakian.
\newblock {Strange quark matter in strong magnetic fields within a confining
  model}.
\newblock {\em Phys. Rev.}, D88(2):025008, 2013.

\bibitem{Sinha:2015bva}
Monika Sinha and Armen Sedrakian.
\newblock {Magnetar superconductivity versus magnetism: neutrino cooling
  processes}.
\newblock {\em Phys. Rev.}, C91(3):035805, 2015.

\bibitem{Huang:2009ue}
Xu-Guang Huang, Mei Huang, Dirk~H. Rischke, and Armen Sedrakian.
\newblock {Anisotropic Hydrodynamics, Bulk Viscosities and R-Modes of Strange
  Quark Stars with Strong Magnetic Fields}.
\newblock {\em Phys. Rev.}, D81:045015, 2010.

\bibitem{Anand:1979uq}
J.~D. Anand, S.~N. Biswas, and M.~Hasan.
\newblock {Magnetization Of Quark Gas In Intense Magnetic Field}.
\newblock {\em J. Phys.}, A12:L235--L237, 1979.

\bibitem{Fulde:1964zz}
Peter Fulde and Richard~A. Ferrell.
\newblock {Superconductivity in a Strong Spin-Exchange Field}.
\newblock {\em Phys. Rev.}, 135:A550--A563, 1964.

\bibitem{Larkin:1964zz}
A.~I. larkin and Y.~N. Ovchinnikov.
\newblock {Nonuniform state of superconductors}.
\newblock {\em Zh. Eksp. Teor. Fiz.}, 47:1136--1146, 1964.
\newblock [Sov. Phys. JETP20,762(1965)].

\bibitem{Alford:2010qf}
Mark~G. Alford and Armen Sedrakian.
\newblock {Color-magnetic flux tubes in quark matter cores of neutron stars}.
\newblock {\em J. Phys.}, G37:075202, 2010.

\bibitem{Feng:2011fj}
Bo~Feng, Efrain~J. Ferrer, and Vivian de~la Incera.
\newblock {Cooper Pair's Magnetic Moment in MCFL Color Superconductivity}.
\newblock {\em Nucl. Phys.}, B853:213--239, 2011.

\bibitem{Ouyed:2005dz}
Rachid Ouyed, Brian Niebergal, Wolfgang Dobler, and Denis Leahy.
\newblock {3-dimensional simulations of the reorganization of a quark star's
  magnetic field as induced by the meissner effect}.
\newblock {\em Astrophys. J.}, 653:558--567, 2006.

\bibitem{Thompson:2006ii}
Todd~A. Thompson.
\newblock {Assessing Millisecond Proto-Magnetars as GRB Central Engines}.
\newblock {\em Rev. Mex. Astron. Astrof. Ser. Conf.27,80(2007)}, 2006.

\end{thebibliography}
\bibliographystyle{unsrt}

\end{document}